\documentclass{rsproca_new}

\usepackage{tikz}
\usetikzlibrary{positioning,calc}
\usepackage[percent]{overpic}
\usepackage{xspace}
\usepackage{bm}
\usepackage{caption}
\usepackage{subcaption}
\usepackage{siunitx}
\usepackage{tabularx}
\usepackage{cleveref}

\crefformat{equation}{(#2#1#3)}
\Crefformat{equation}{Equation~(#2#1#3)}
\crefrangeformat{equation}{(#3#1#4)--(#5#2#6)}
\Crefrangeformat{equation}{Equations (#3#1#4)--(#5#2#6)}

\crefformat{figure}{#2Figure~#1#3}
\Crefformat{figure}{#2Figure~#1#3}

\crefformat{table}{#2Table~#1#3}
\Crefformat{table}{#2Table~#1#3}

\crefformat{section}{#2Section~#1#3}
\Crefformat{section}{#2Section~#1#3}

\newcommand{\Hinf}{$\mathcal{H}_\infty$}
\newcommand{\Htwo}{$\mc{H}_2$}
\newcommand{\ie}[0]{{i.e.\ }\xspace}
\newcommand{\mc}[1]{\ensuremath{\mathcal{#1}}}

\newcommand{\mbc}[1]{\ensuremath{\boldsymbol{\mathcal{#1}}}}
\DeclareMathAlphabet{\mbf}{OT1}{ptm}{b}{n}
\newcommand{\mbs}[1]{\ensuremath{\boldsymbol{#1}}}

\newcommand{\mbfbar}[1]{\ensuremath{\bar{\mbf{#1}}}}
\newcommand{\mbfhat}[1]{\ensuremath{\hat{\mbf{#1}}}}
\newcommand{\mbftilde}[1]{\ensuremath{\tilde{\mbf{#1}}}}

\newcommand{\mbshat}[1]{\ensuremath{\hat{\boldsymbol{#1}}}}
\newcommand{\mbstilde}[1]{\ensuremath{\tilde{\boldsymbol{#1}}}}
\newcommand{\trans}{{\ensuremath{\mathsf{T}}}}
\newcommand{\frob}{{\ensuremath{\mathsf{F}}}}
\newcommand{\trace}{{\ensuremath{\mathrm{tr}}}}
\newcommand{\He}[1]{{\ensuremath{\mathrm{He}\!\left\{#1\right\}}}}
\newcommand{\bma}[1]{\left[\begin{array}{#1}}
\newcommand{\ema}{\end{array}\right]}

\definecolor{oiorange}{rgb}{0.90, 0.60, 0.00}
\definecolor{oiskyblue}{rgb}{0.35, 0.70, 0.90}
\definecolor{oibluishgreen}{rgb}{0.00, 0.60, 0.50}
\definecolor{oiyellow}{rgb}{0.95, 0.90, 0.25}
\definecolor{oiblue}{rgb}{0.00, 0.45, 0.70}
\definecolor{oivermillion}{rgb}{0.80, 0.40, 0.00}
\definecolor{oireddishpurple}{rgb}{0.80, 0.60, 0.70}

\jname{rspa}
\Journal{Proc R Soc A\ }

\begin{document}

\title{System Norm Regularization Methods for Koopman Operator Approximation}

\author{Steven Dahdah$^{1}$ and James R. Forbes$^{1}$}

\address{$^{1}$Department of Mechanical Engineering, McGill University, Montreal
QC H3A~0C3, Canada}

\subject{applied mathematics, mathematical modelling, robotics, control}


\corres{Steven Dahdah\\\email{steven.dahdah@mail.mcgill.ca}}

\begin{abstract}
    Approximating the Koopman operator from data is numerically challenging when
    many lifting functions are considered. Even low-dimensional systems can
    yield unstable or ill-conditioned results in a high-dimensional lifted
    space. In this paper, Extended Dynamic Mode Decomposition (DMD) and DMD with
    control, two methods for approximating the Koopman operator, are
    reformulated as convex optimization problems with linear matrix inequality
    constraints.
    Asymptotic stability constraints and system norm regularizers are then
    incorporated as methods to improve the numerical conditioning of the Koopman
    operator. Specifically, the \Hinf{}~norm is used to penalize the
    input-output gain of the Koopman system. Weighting functions are then
    applied to penalize the system gain at specific frequencies.
    These constraints and regularizers introduce bilinear matrix inequality
    constraints to the regression problem, which are handled by solving a
    sequence of convex optimization problems.
    Experimental results using data from an aircraft fatigue structural test rig
    and a soft robot arm highlight the advantages of the proposed regression
    methods.
\end{abstract}

\begin{fmtext}
    \section{Introduction}\label{sec:introduction}
    Koopman operator theory~\cite{koopman_hamiltonian_1931, mezic_2019_spectrum,
    budisic_applied_2012, mauroy_2020_koopman} allows a nonlinear system to be
    exactly represented as a linear system in terms of an infinite set of
    \textit{lifting functions}.
    The \textit{Koopman operator} advances each of these lifting function to the
    next timestep.
    Thanks to recent theoretical developments~\cite{mezic_2019_spectrum,
    budisic_applied_2012, mauroy_2020_koopman} and the widespread availability of
    computational resources, there has been a recent resurgence of interest in using
    data-driven methods to approximate the Koopman operator.
    The Koopman operator defines a linear state-space system in the chosen lifted
    space, making it convenient for control system design.
    Koopman models have been paired with a wide variety of existing linear
    optimal control techniques~\cite{korda_2018_linear, otto_2021_koopman,
    abraham_active_2019, mamakoukas_local_2019, bruder_modeling_2019,
    uchida_2021_data-driven} with great success.
\end{fmtext}

\maketitle

\tikzstyle{overview} = [
    draw=gray,
    rounded corners=0.1cm,
    line width=1.5pt,
    minimum height=14cm,
    inner sep=0.1cm,
    anchor=north west,
    align=center,
]
\begin{figure}[ht]
    \centering
    \resizebox{\textwidth}{!}{\begin{tikzpicture}
        \node[
            overview,
            minimum width=4.5cm,
            label={north:(a) Collect data snapshots},
        ] (step1) at (0, 0) {%
            \begin{overpic}[width=3cm]{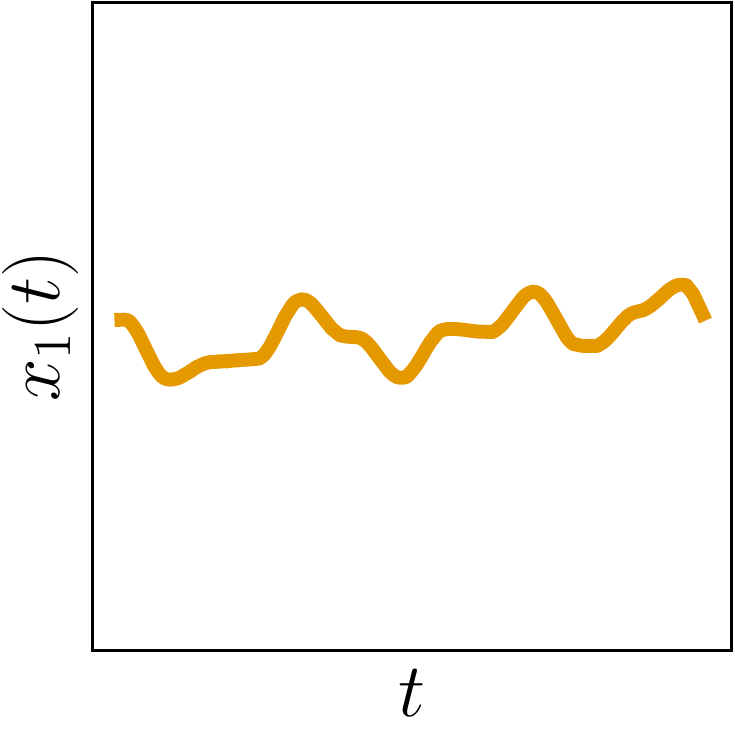}
               \put(-10, -10){\includegraphics[width=3cm]{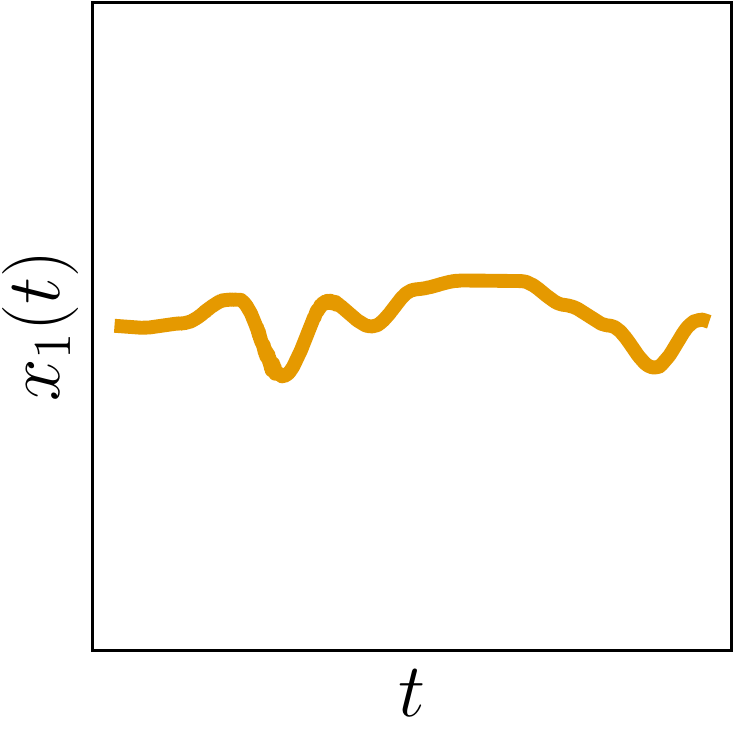}}
               \put(-20, -20){\includegraphics[width=3cm]{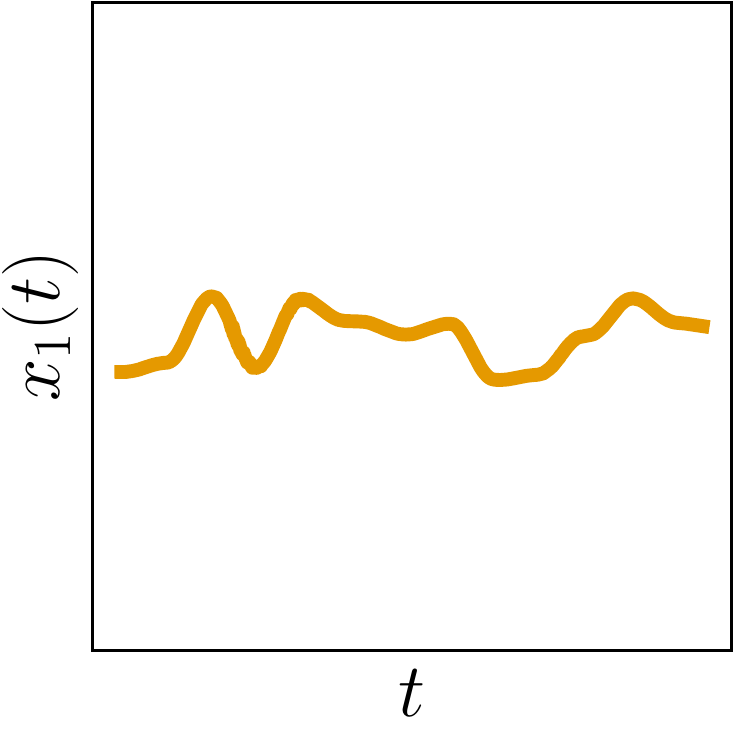}}
            \end{overpic}\\[8ex]
            \begin{overpic}[width=3cm]{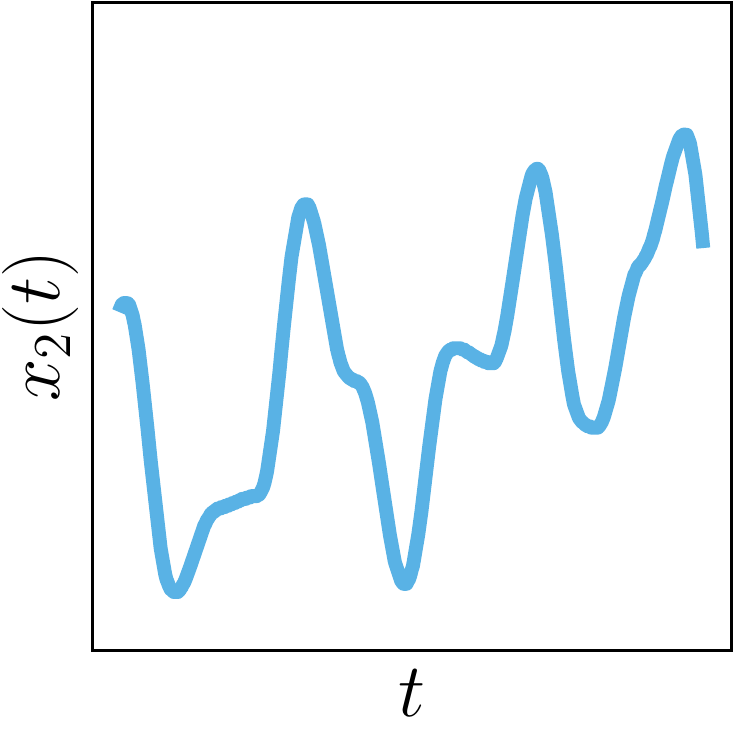}
               \put(-10, -10){\includegraphics[width=3cm]{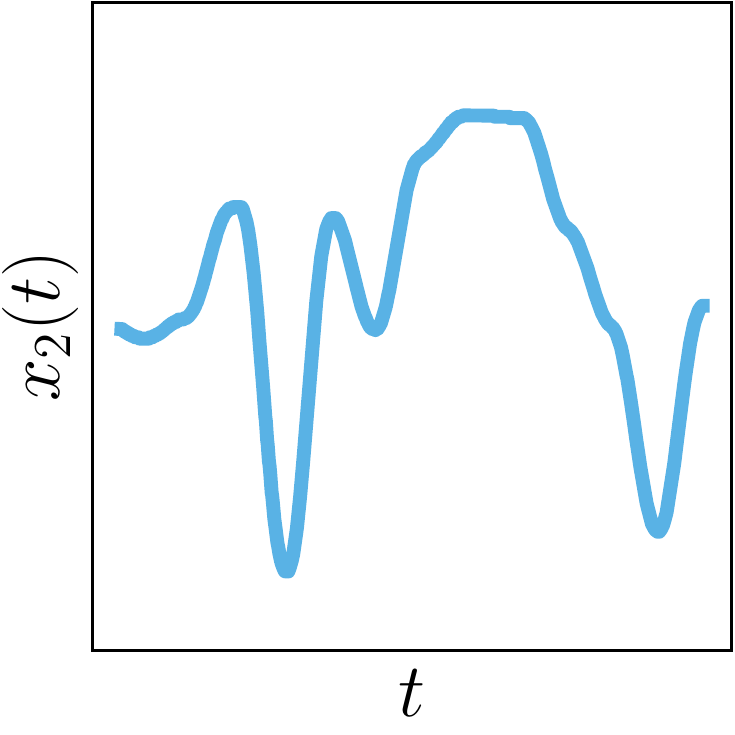}}
               \put(-20, -20){\includegraphics[width=3cm]{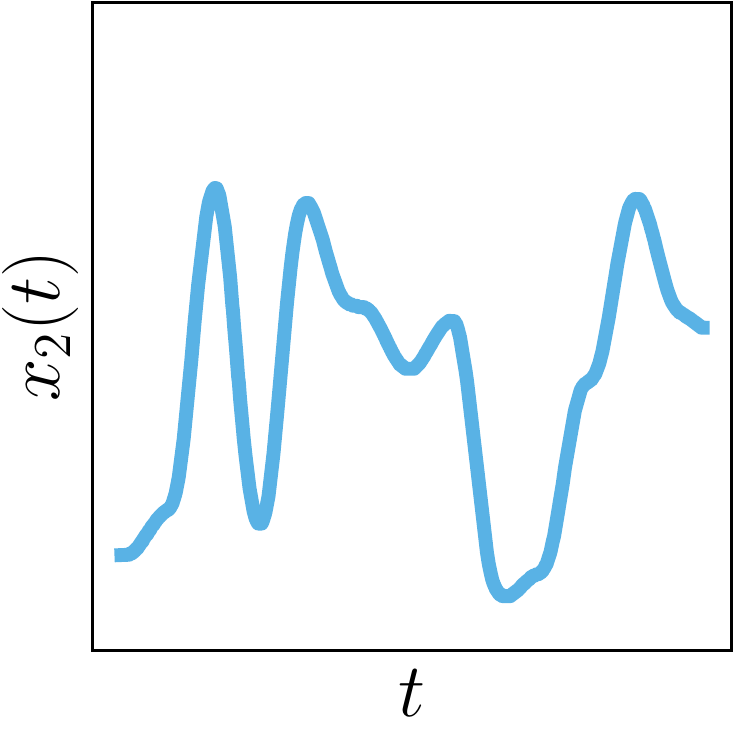}}
            \end{overpic}\\[8ex]
            \begin{overpic}[width=3cm]{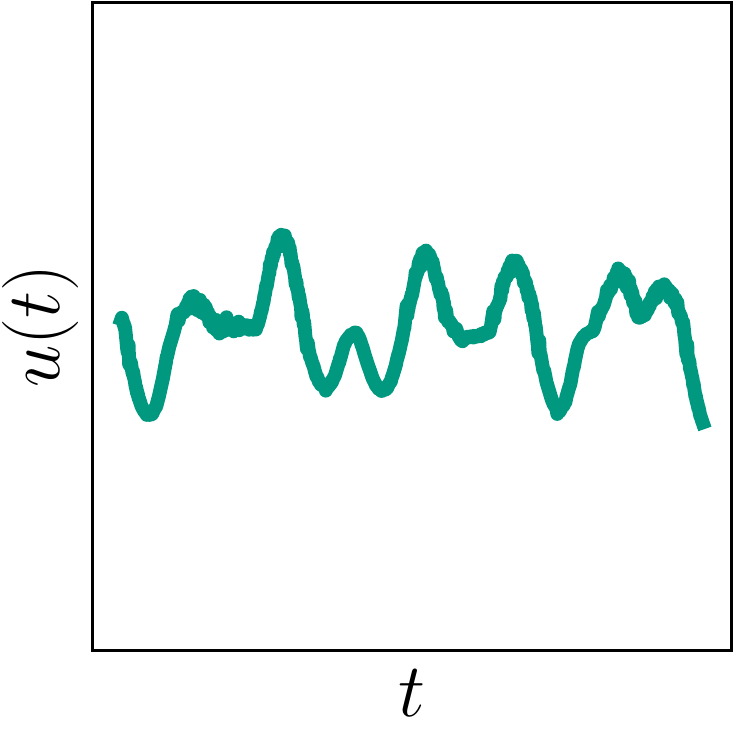}
               \put(-10, -10){\includegraphics[width=3cm]{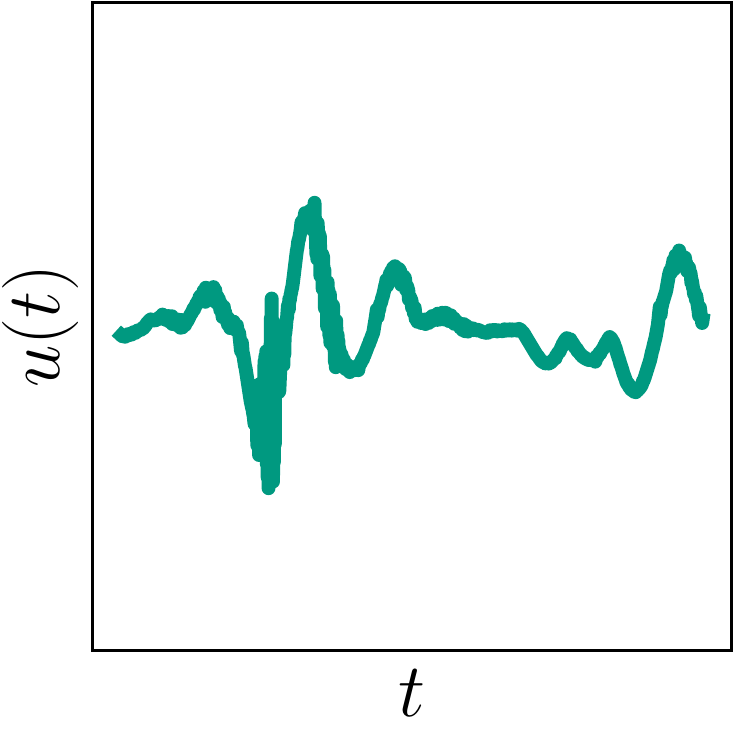}}
               \put(-20, -20){\includegraphics[width=3cm]{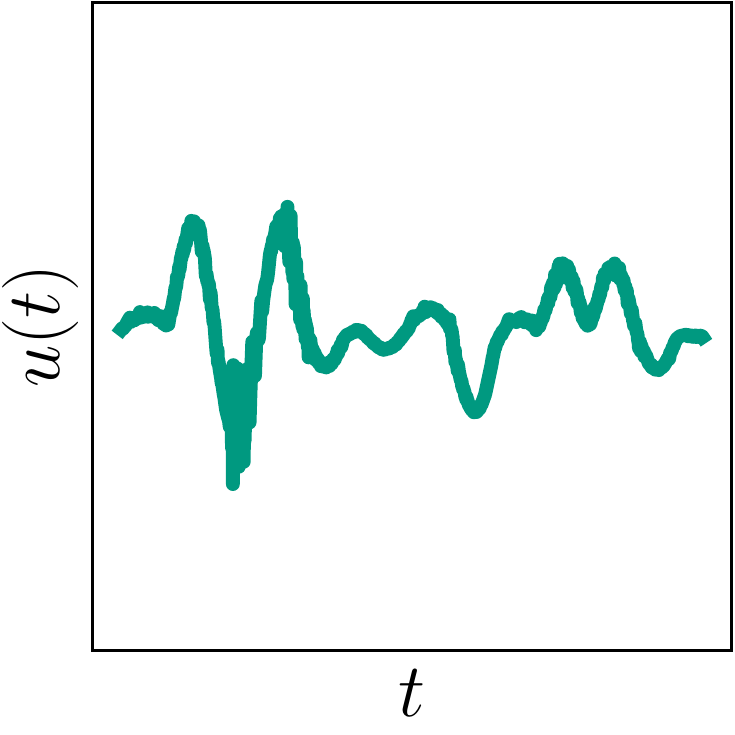}}
            \end{overpic}
        };
        \node[
            overview,
            right=0.25cm of step1,
            label={north:(b) Select lifting functions},
        ] (step2) {%
            \includegraphics[width=4cm]{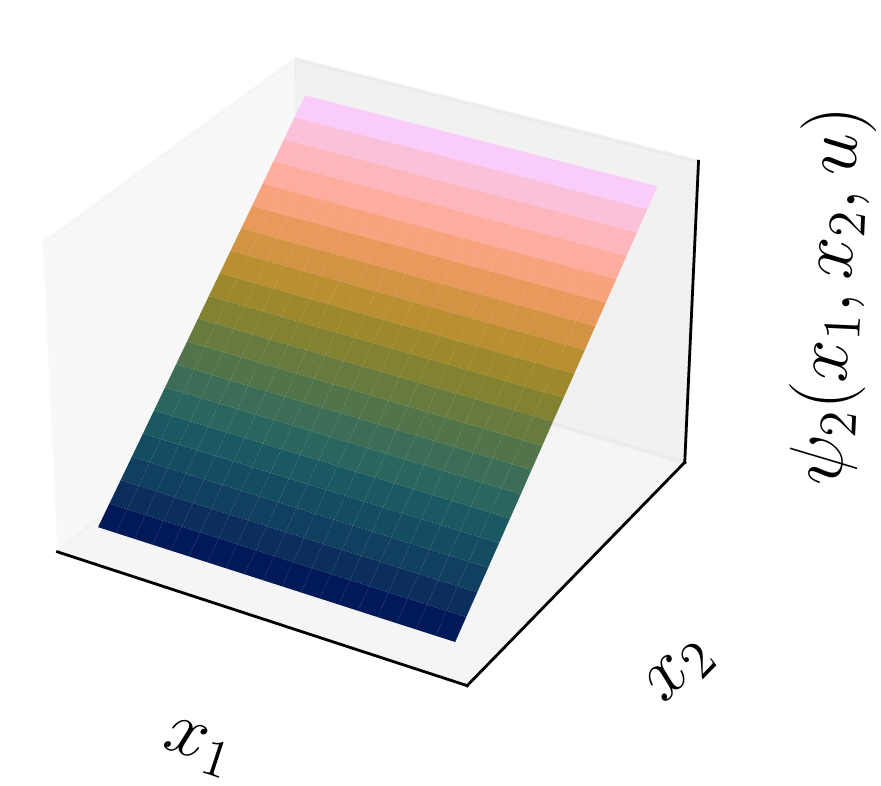}\\
            $\psi_2(x_1, x_2, u) = x_2$\\
            \includegraphics[width=4cm]{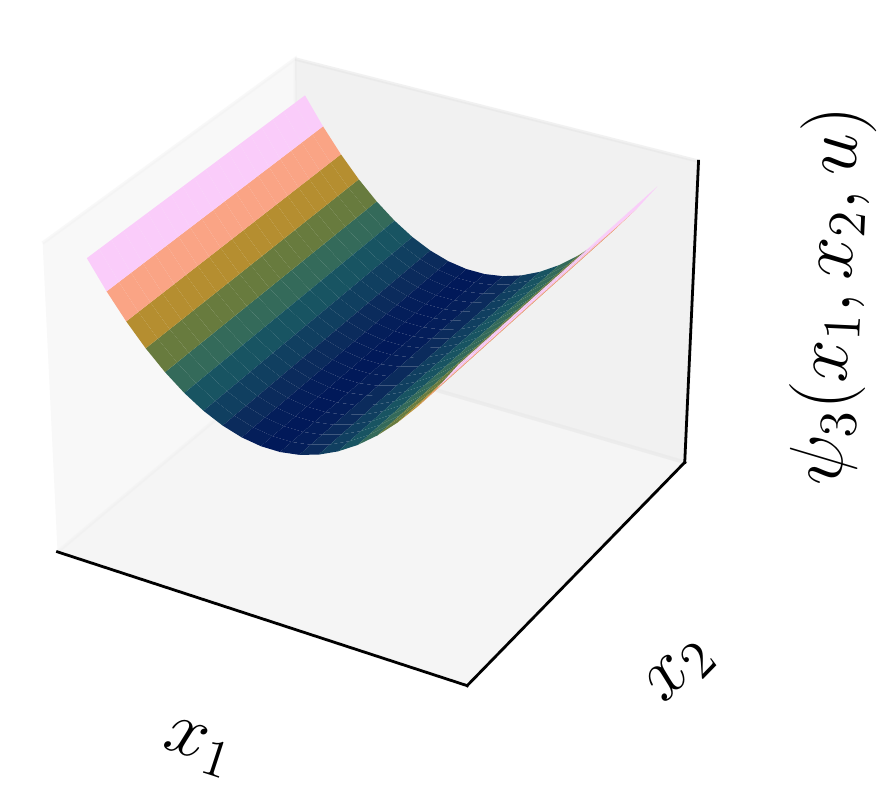}\\
            $\psi_3(x_1, x_2, u) = x_1^2$\\
            \includegraphics[width=4cm]{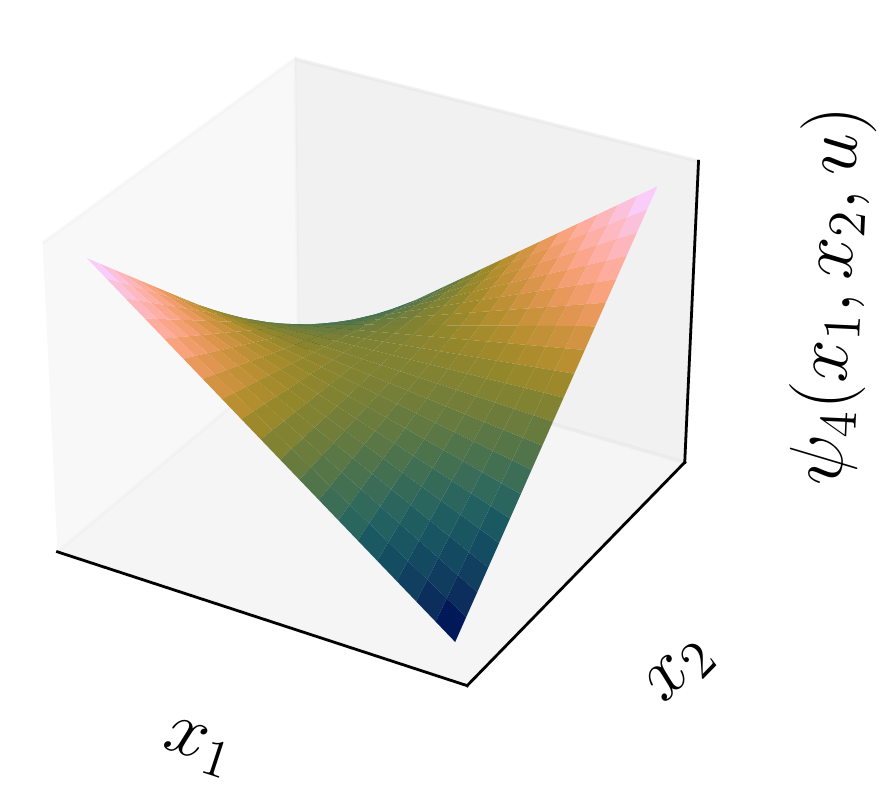}\\
            $\psi_4(x_1, x_2, u) = x_1 x_2$
        };
        \node[
            overview,
            right=0.25cm of step2,
            label={north:(c) Lift and arrange snapshots},
        ] (step3) {%
            \begin{overpic}[width=2.5cm]{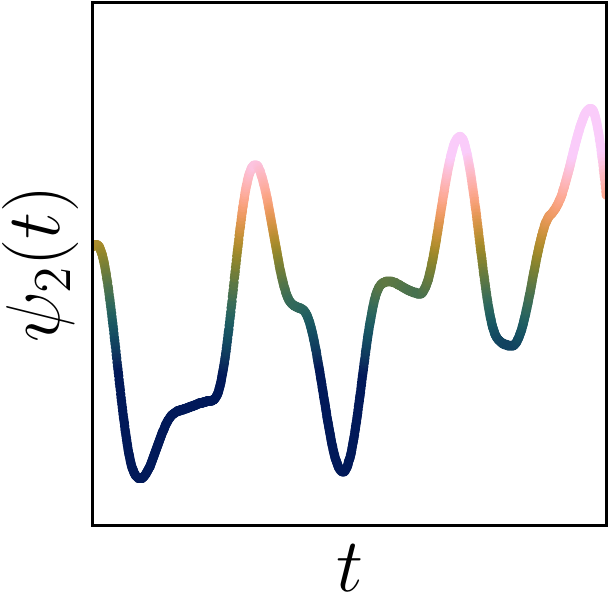}
               \put(-10, -10){\includegraphics[width=2.5cm]{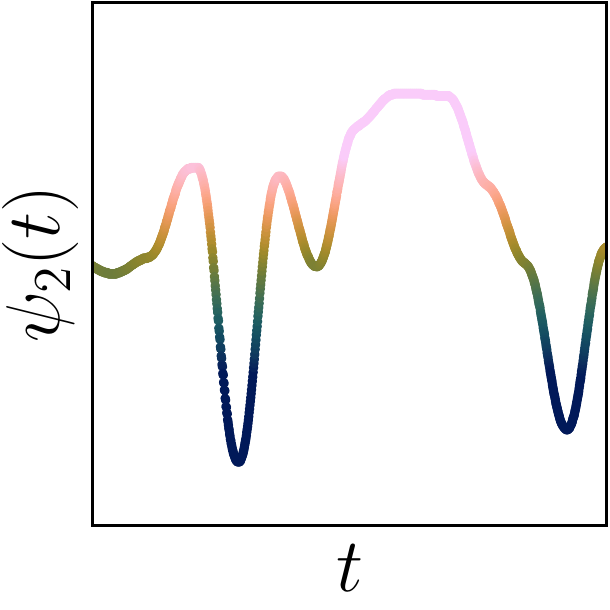}}
               \put(-20, -20){\includegraphics[width=2.5cm]{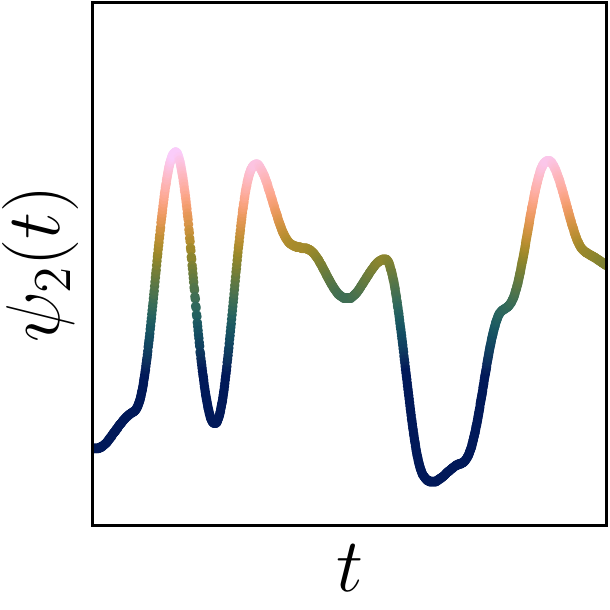}}
            \end{overpic}\\[5ex]
            \begin{overpic}[width=2.5cm]{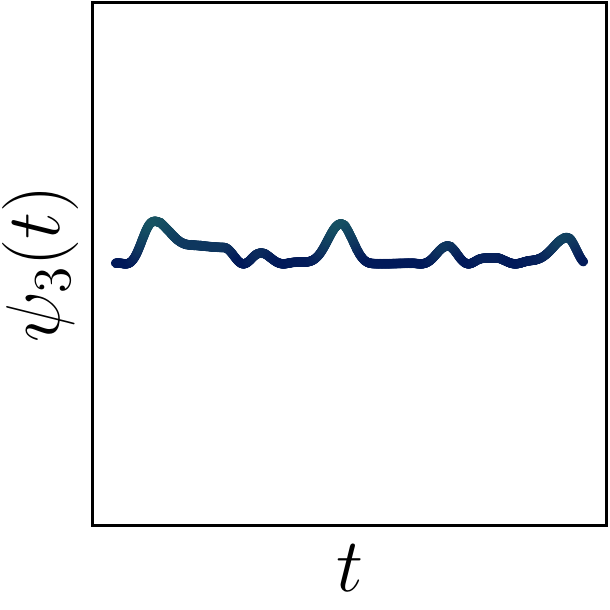}
               \put(-10, -10){\includegraphics[width=2.5cm]{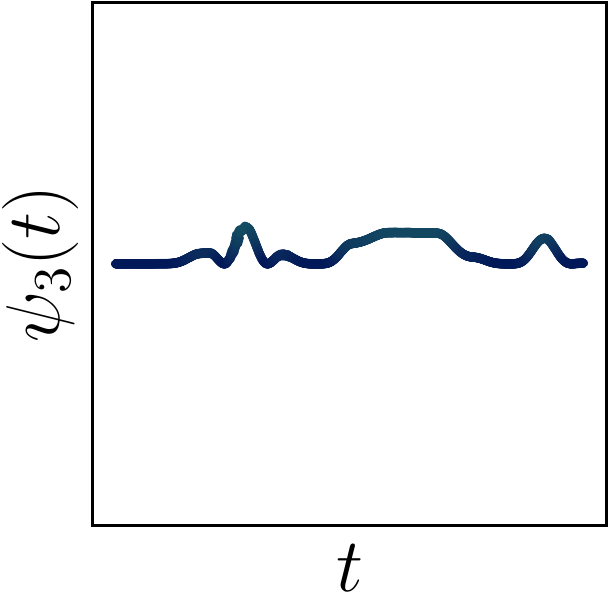}}
               \put(-20, -20){\includegraphics[width=2.5cm]{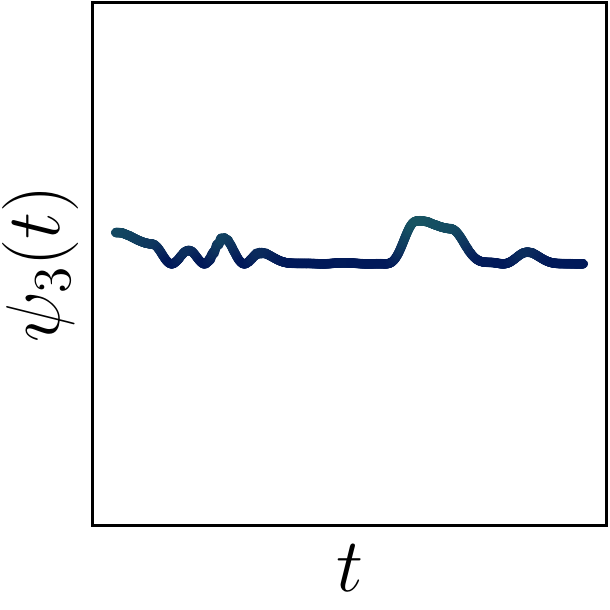}}
            \end{overpic}\\[5ex]
            \begin{overpic}[width=2.5cm]{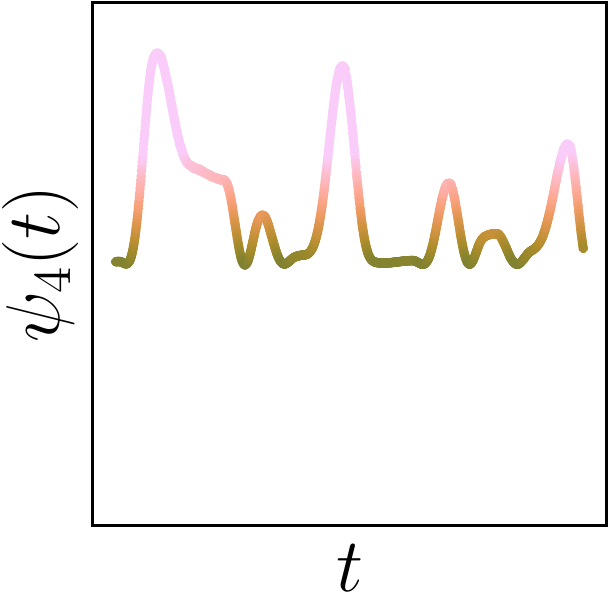}
               \put(-10, -10){\includegraphics[width=2.5cm]{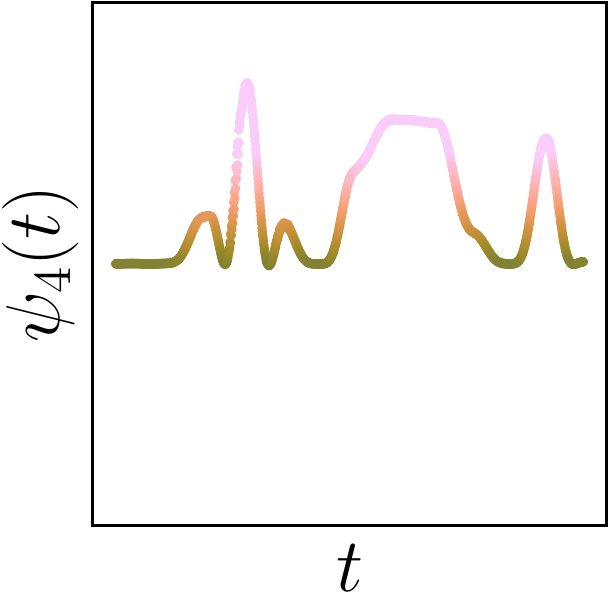}}
               \put(-20, -20){\includegraphics[width=2.5cm]{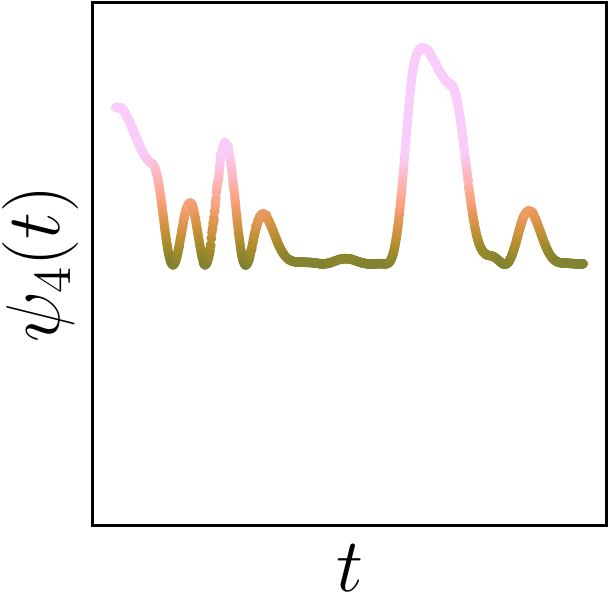}}
            \end{overpic}\\[3ex]
            \begin{tikzpicture}
                \draw[-latex, line width=2pt] (0cm, 0cm) -- (0cm, -0.8cm);
            \end{tikzpicture}\\
            $\begin{bmatrix}
                | & | & & | \\
                \mbs{\psi}_0 &
                    \mbs{\psi}_1 &
                    \cdots &
                    \mbs{\psi}_{q-1} \\
                | & | & & |
            \end{bmatrix} = \mbs{\Psi}$
            \\[2ex]
            $\begin{bmatrix}
                | & | & & | \\
                \mbs{\vartheta}_1 &
                    \mbs{\vartheta}_2 &
                    \cdots &
                    \mbs{\vartheta}_q \\
                | & | & & |
            \end{bmatrix} = \mbs{\Theta}_+$
        };
        \node[
            overview,
            right=0.25cm of step3,
            label={north:(d) Approximate Koopman operator},
        ] (step4) {%
            \textit{Extended DMD (\cref{sec:edmd}):}\\
            $\mbf{U}^\ast = \arg \underset{\mbf{U}}{\min}\|\mbs{\Theta}_+ - \mbf{U}\mbs{\Psi}\|_\frob^2$
            \\[4ex]
            \textit{Asymptotic stability constraint (\cref{sec:srconst}):}\\
            $\mbf{U}^\ast = \arg \underset{\mbf{U}}{\min}\|\mbs{\Theta}_+ - \mbf{U}\mbs{\Psi}\|_\frob^2$ \\
            $\mathrm{s.t.}\ \color{oiblue}{\bar{\lambda}(\mbf{A}) < \bar{\rho}}$\\[1ex]
            \includegraphics[width=4.4cm]{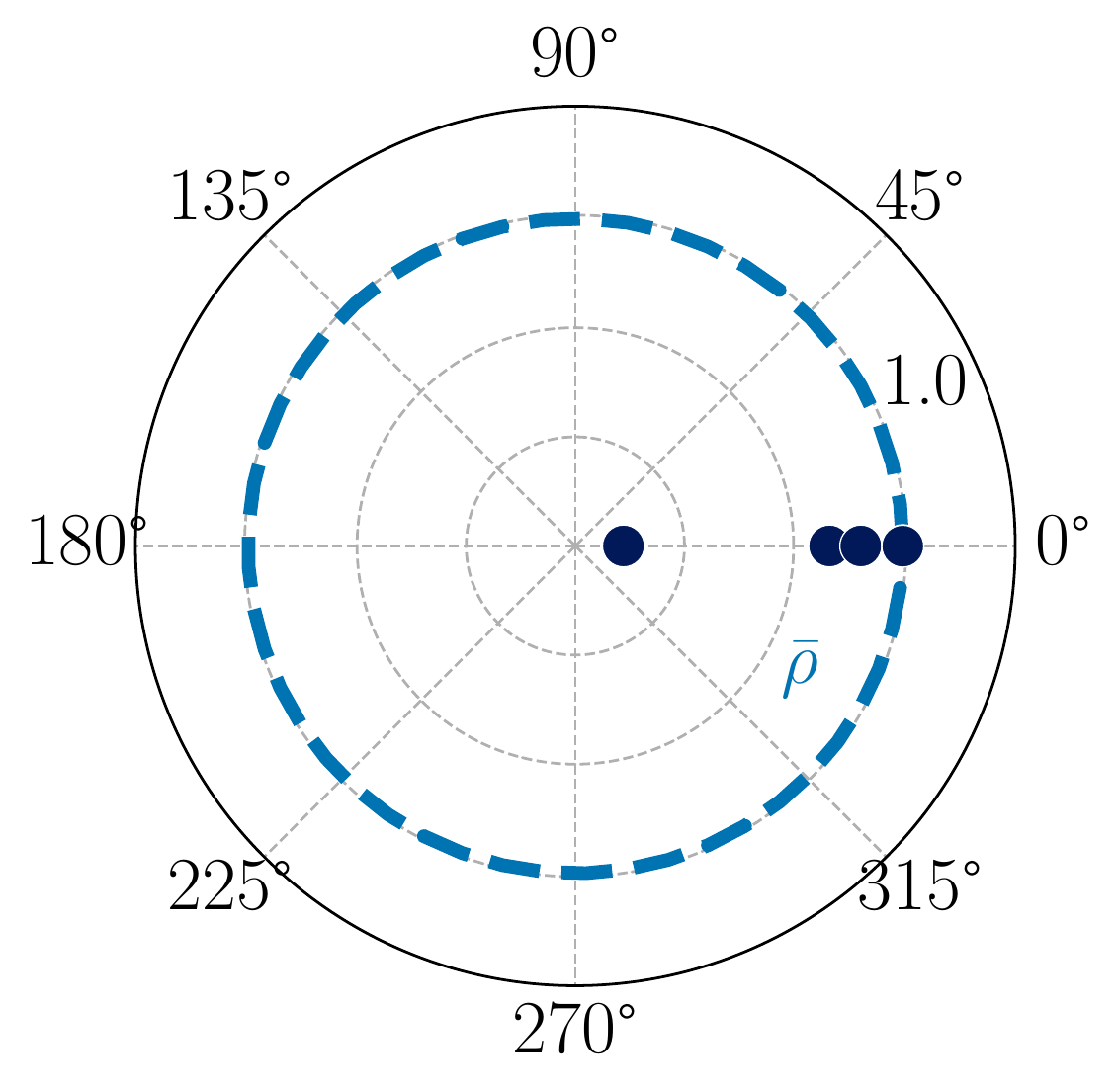}
            \\[4ex]
            \textit{\Hinf{}~norm regularizer (\cref{sec:hinf}):}\\
            $\mbf{U}^\ast = \arg \underset{\mbf{U}}{\min}\|\mbs{\Theta}_+ - \mbf{U}\mbs{\Psi}\|_\frob^2 + \color{oivermillion}{\beta \|\mbc{G}\|_\infty}$\\[1ex]
            \includegraphics[width=4.4cm]{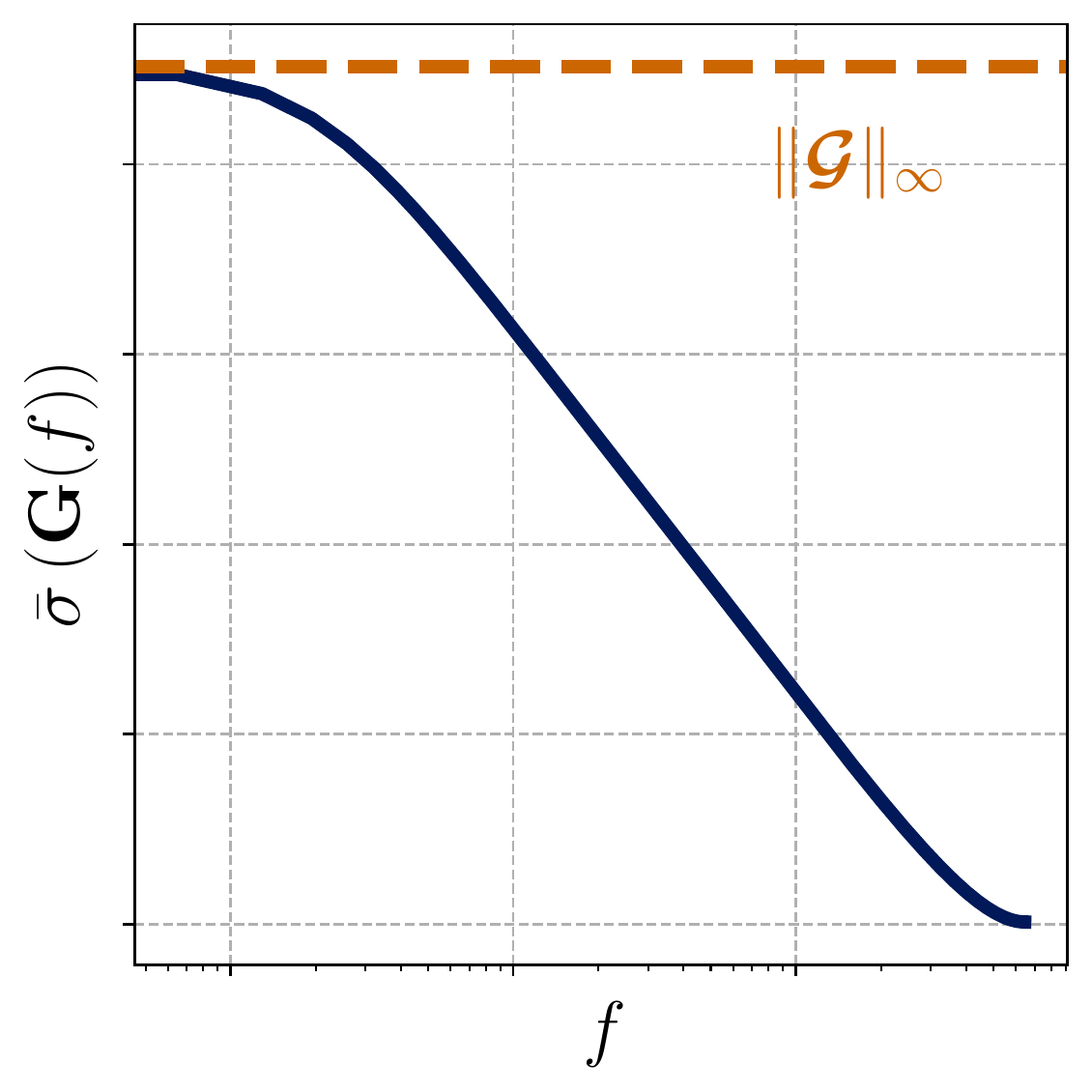}
        };
    \end{tikzpicture}}
    \caption{Overview of the data-driven Koopman workflow, including the role of
    the proposed regression methods. (a)~First, data is collected from the
    system to be identified. (b)~Next, a set of lifting functions is chosen.
    (c)~The lifting functions are then applied to the data and snapshot matrices
    are formed. (d)~Finally, one of several regression methods is used to
    approximate the Koopman matrix. The proposed regression methods seek to
    address the numerical problems often present in this step by viewing it as
    an optimization over discrete-time linear systems.}\label{fig:overview}
\end{figure}

To use the Koopman representation in practical applications, a
finite-dimensional approximation of the infinite-dimensional Koopman operator
must be found.
First, a finite set of lifting functions is selected. These functions are often
hand-picked based on known dynamics~\cite{abraham_active_2019,
mamakoukas_local_2019}, or are combinations of sinusoids, polynomials, and other
basis functions~\cite{bruder_modeling_2019, abraham_model-based_2017}. Time
delay embeddings are also popular~\cite{korda_2018_linear,
bruder_modeling_2019}.
However, there is no universally agreed-upon method for selecting lifting
functions.
Given a set of lifting functions, linear regression is used to find the matrix
approximation of the Koopman operator, also called a \textit{Koopman
matrix}~\cite{kutz_dynamic_2016, otto_2021_koopman}.

Unfortunately, the regression problem associated with finding an approximate
Koopman operator is numerically challenging, as complex lifting function choices
can yield unstable or ill-conditioned Koopman models for stable
systems~\cite{mamakoukas_2020_learning}.
Regularization techniques play a crucial role in obtaining usable Koopman models
for prediction and control applications.
Standard regularization techniques like Tikhonov
regularization~\cite{tikhonov_1995_numerical} or the
lasso~\cite{tibshirani_regression_1996} are often used to
promote well-conditioned Koopman matrices.
These regularization techniques penalize different matrix norms of the Koopman
matrix, without considering the fact that the Koopman matrix defines a
discrete-time linear system with input, state, and output.
While these methods may indirectly promote asymptotic stability in this Koopman
system, their success is highly dependent on the regularization coefficient
used. Furthermore, they do not consider the input-output gain of the Koopman
system.
This paper takes a systems view of the Koopman matrix regression problem,
proposing regression methods that constrain the asymptotic stability of the
Koopman system and regularize the regression problem by penalizing its
input-output gain, as represented by its \Hinf{}~norm.
To accomplish this, the \textit{Extended Dynamic Mode Decomposition}
(EDMD)~\cite{williams_data-driven_2015} and \textit{Dynamic Mode Decomposition
with control} (DMDc)~\cite{proctor_2014_dynamic} methods are reformulated as
convex optimization problems with linear matrix inequality (LMI) constraints.
Regularizers and additional constraints are incorporated in a modular fashion as
either LMI constraints, or bilinear matrix inequality (BMI) constraints.
The data-driven Koopman workflow and the proposed regression methods
are summarized in~\cref{fig:overview}.

\subsection{Related work}
Convex optimization and LMIs have previously been used to synthesize controllers
for Koopman models~\cite{uchida_2021_data-driven}, but they have not yet
been leveraged to regularize the Koopman matrix regression problem.
A related optimization problem is posed in~\cite{sznaier_2021_convex}, where
both the Koopman matrix and lifting functions are treated as unknowns. While
this problem is NP-hard, a convex relaxation allows both to be found by solving
two semidefinite programs.
The \Hinf{}~norm of the Koopman operator has previously been considered
in~\cite{hara_2020_learning}, however it is in the form of a hard constraint on
the system's dissipativity, rather than as a regularizer in the cost function.

In~\cite{lacy_subspace_2002}, the problem of learning Lyapunov stable and
asymptotically stable linear systems is explored in the context of subspace
identification, where Lyapunov inequalities are used to enforce the
corresponding stability conditions.
The related problem of learning positive real and strictly positive real systems
using constrained subspace identification is discussed
in~\cite{hoagg_first-order-hold_2004}.
A convex relaxation of the Lyapunov inequality is considered
in~\cite{boots_2008_constraint}, where linear constraints are added
incrementally to enforce the Lyapunov stability of a system.
In~\cite{mamakoukas_2020_learning}, a gradient-descent method called
\textit{SOC}~\cite{mamakoukas_2020_memory} is applied to find locally optimal
Lyapunov stable or asymptotically stable Koopman matrices. The method relies on
a parameterization of the Koopman matrix that guarantees Lyapunov stability or
asymptotic stability~\cite{gillis_2020_note}.
While addressing the asymptotic stability problem, this formulation lacks the
modularity of the proposed approach.

\subsection{Contribution}
The core contributions of this paper are solving the EDMD and DMDc problems
with asymptotic stability constraints and with system norm regularizers. Of
particular focus is the use of the \Hinf{}~norm as a regularizer, which
penalizes the worst-case gain of the Koopman system over all frequencies.
The BMI formulation of the EDMD problem with asymptotic stability constraints
and \Hinf{}~norm regularization were previously explored by the authors
in~\cite{dahdah_2021_linear}. LMI formulations for Tikhonov regularization,
matrix two-norm regularization, and nuclear norm regularization were also
presented in~\cite{dahdah_2021_linear}.
This paper expands on~\cite{dahdah_2021_linear} to include an LMI formulation of
the DMDc problem, and discusses the corresponding asymptotic stability
constraint and \Hinf{}~norm regularizer. As with EDMD, these modifications add
BMI constraints to the DMDc problem. Furthermore, weighted \Hinf{}~norm
regularization is explored, which allows the Koopman system's gain to be
penalized in a specific frequency band, where experimental measurements may be
less reliable, or where system dynamics may be irrelevant.
Finally, the proposed regression methods are evaluated using two experimental
datasets, one from a fatigue structural testing platform, and the other from a
soft robot arm.
The significance of this work is the use of a system norm to regularize the
Koopman regression problem, which is viewed as a regression problem over
discrete-time linear systems, resulting in a numerically better conditioned
data-driven model.

\section{Background}\label{sec:background}
\subsection{Koopman operator theory}
Consider the discrete-time nonlinear process
\begin{equation}
    \mbf{x}_{k+1} = \mbf{f}(\mbf{x}_{k}),%
    \label{eq:dyn_noin}
\end{equation}
where
${\mbf{x}_{k} \in \mc{M}}$
evolves on a manifold
${\mc{M} \subseteq \mathbb{R}^{m \times 1}}$.
Let
${\psi: \mc{M} \to \mathbb{R}}$
be a \textit{lifting function}.
Any scalar function of the state $\mbf{x}_{k}$ qualifies as a lifting function.
The lifting functions therefore form an infinite-dimensional Hilbert space
$\mc{H}$.
The \textit{Koopman operator}
${\mc{U}: \mc{H} \to \mc{H}}$
is a linear operator that advances all scalar-valued lifting functions
${\psi \in \mc{H}}$
in time by one timestep. That is~\cite[\S3.2]{kutz_dynamic_2016},
\begin{equation}
    (\mc{U} \psi)(\cdot) = (\psi \circ \mbf{f})(\cdot).%
    \label{eq:koopman_def_noin}
\end{equation}
Using~\cref{eq:koopman_def_noin}, the dynamics of~\cref{eq:dyn_noin} can be
rewritten linearly in terms of $\psi$ as
\begin{equation}
    \psi(\mbf{x}_{k+1}) = (\mc{U} \psi)(\mbf{x}_{k}). \label{eq:koopman_dyn}
\end{equation}
In finite dimensions,~\cref{eq:koopman_dyn} is approximated by
\begin{equation}
    \mbs{\psi}(\mbf{x}_{k+1}) = \mbf{U} \mbs{\psi}(\mbf{x}_{k}) + \mbf{r}_k,%
    \label{eq:koopman_approx_noin}
\end{equation}
where
${\mbs{\psi}: \mc{M} \to \mathbb{R}^{p \times 1}}$,
${\mbf{U} \in \mathbb{R}^{p \times p}}$,
and $\mbf{r}_k$ is the residual error.
Each element of the \textit{vector-valued lifting function} $\mbs{\psi}$ is
a lifting function in $\mc{H}$.
The \textit{Koopman matrix} $\mbf{U}$ is a matrix approximation of the Koopman
operator.

\subsection{Koopman operator theory with inputs}
If a discrete-time nonlinear process with exogenous inputs is considered, the
definitions of the lifting functions and Koopman operator must be modified.
Consider
\begin{equation}
    \mbf{x}_{k+1} = \mbf{f}(\mbf{x}_{k}, \mbf{u}_k), \label{eq:dyn_in}
\end{equation}
where
${\mbf{x}_{k} \in \mc{M} \subseteq \mathbb{R}^{m \times 1}}$
and
${\mbf{u}_{k} \in \mc{N} \subseteq \mathbb{R}^{n \times 1}}$.
In this case, the lifting functions become
${\psi: \mc{M} \times \mc{N} \to \mathbb{R}}$
and the Koopman operator
${\mc{U}: \mc{H} \to \mc{H}}$
is instead defined so that
\begin{equation}
    (\mc{U} \psi)(\mbf{x}_{k}, \mbf{u}_{k})
    = \psi(\mbf{f}(\mbf{x}_{k}, \mbf{u}_k), \star),
    \label{eq:koop-def-in}
\end{equation}
where
${\star = \mbf{u}_k}$
if the input has state-dependent dynamics, or
${\star = \mbf{0}}$
if the input has no dynamics~\cite[\S6.5]{kutz_dynamic_2016}. If the input is
computed by a controller, it is often considered to have state-dependent
dynamics.
Let the vector-valued lifting function
${\mbs{\psi}: \mc{M} \times \mc{N} \to \mathbb{R}^{p \times 1}}$
be partitioned as
\begin{equation}
    \mbs{\psi}(\mbf{x}_k, \mbf{u}_k) = \begin{bmatrix}
        \mbs{\vartheta}(\mbf{x}_k) \\
        \mbs{\upsilon}(\mbf{x}_k, \mbf{u}_k)
    \end{bmatrix},
\end{equation}
where
${\mbs{\vartheta}: \mc{M} \to \mathbb{R}^{p_\vartheta \times 1}}$,
${\mbs{\upsilon}: \mc{M} \times \mc{N} \to \mathbb{R}^{p_\upsilon \times 1}}$,
and
${p_\vartheta + p_\upsilon = p}$.
When the input is exogenous,~\cref{eq:koop-def-in} has
the form~\cite[\S6.5.1]{kutz_dynamic_2016}
\begin{equation}
    \mbs{\vartheta}(\mbf{x}_{k+1}) \\
    =
    \mbf{U}
    \mbs{\psi}(\mbf{x}_{k}, \mbf{u}_{k})
    + \mbf{r}_k,%
    \label{eq:U_part}
\end{equation}
where ${\mbf{U} = \begin{bmatrix} \mbf{A} & \mbf{B} \end{bmatrix}}$.
Expanding~\cref{eq:U_part} yields the familiar linear state-space form,
\begin{equation}
    \mbs{\vartheta}(\mbf{x}_{k+1})
    =
    \mbf{A} \mbs{\vartheta}(\mbf{x}_{k})
    + \mbf{B} \mbs{\upsilon}(\mbf{x}_k, \mbf{u}_k)
    + \mbf{r}_k.
    \label{eq:dyn_approx_ss_in}
\end{equation}

When identifying a Koopman model for control, the input is often left unlifted,
that is,
${\mbs{\upsilon}(\mbf{x}_k, \mbf{u}_k) = \mbf{u}_k}$~\cite{korda_2018_linear}.
However, recent work demonstrates that this choice of lifting functions is
insufficient for describing control affine systems, which are ubiquitous in
real-world applications~\cite{bruder_2021_advantages}.
An alternative choice of input-dependent lifting functions proposed
in~\cite{bruder_2021_advantages} is
\begin{equation}
    \mbs{\upsilon}(\mbf{x}_k, \mbf{u}_k)
    =
    \begin{bmatrix}
        \mbf{u}_k \otimes \mbs{\vartheta}(\mbf{x}_k) \\
        \mbf{u}_k
    \end{bmatrix},
\end{equation}
where $\otimes$ denotes the Kronecker product.
These bilinear input-dependent lifting functions are capable of representing all
control affine systems, and therefore present an interesting alternative to
leaving the input unlifted~\cite{bruder_2021_advantages}.
%
%

\subsection{Approximating the Koopman operator from data}
To approximate the Koopman matrix from a dataset
${\mc{D} = {\{\mbf{x}_k, \mbf{u}_k\}}_{k=0}^q}$,
consider the lifted snapshot matrices
\begin{align}
    \mbs{\Psi} &= \begin{bmatrix}
        \mbs{\psi}_{0} & \mbs{\psi}_{1} & \cdots & \mbs{\psi}_{q-1}
    \end{bmatrix} \in \mathbb{R}^{p \times q}, \\
    \mbs{\Theta}_+ &= \begin{bmatrix}
        \mbs{\vartheta}_{1} & \mbs{\vartheta}_{2} & \cdots & \mbs{\vartheta}_{q}
    \end{bmatrix} \in \mathbb{R}^{p_\vartheta \times q}, \label{eq:Theta}
\end{align}
where
${\mbs{\psi}_k = \mbs{\psi}(\mbf{x}_k, \mbf{u}_k)}$
and
${\mbs{\vartheta}_k = \mbs{\vartheta}(\mbf{x}_k)}$.
The Koopman matrix that minimizes
\begin{equation}
    J(\mbf{U}) = \|\mbs{\Theta}_+ - \mbf{U} \mbs{\Psi}\|_\frob^2
    \label{eq:koopman-cost}
\end{equation}
is~\cite[\S1.2.1]{kutz_dynamic_2016}
\begin{equation}
    \mbf{U} = \mbs{\Theta}_+ \mbs{\Psi}^\dagger,
    \label{eq:koopman-soln}
\end{equation}
where ${(\cdot)}^\dagger$ denotes the Moore-Penrose pseudoinverse.
%

\section{Reformulating the Koopman Operator Regression Problem}\label{sec:edmd}
\subsection{Extended DMD}
The direct least-squares method of approximating the Koopman operator
in~\cref{eq:koopman-soln} is fraught with numerical and performance issues.
Namely, computing the pseudoinverse of $\mbs{\Psi}$ is costly when the dataset
contains many snapshots. Extended Dynamic Mode Decomposition
(EDMD)~\cite{williams_data-driven_2015} reduces the dimension of the pseudoinverse
required to compute~\cref{eq:koopman-soln}
when the number of snapshots is much larger than the dimension of the
lifted state (\ie{}, ${p \ll q}$)~\cite[\S10.3]{kutz_dynamic_2016}.

Extended DMD consists of computing~\cref{eq:koopman-soln} using
\begin{equation}
    \mbf{U}
    =
    \mbs{\Theta}_+
    \left(
        \mbs{\Psi}^\trans
        \mbs{\Psi}^{\trans^\dagger}
    \right)
    \mbs{\Psi}^\dagger
    =
    \left(\mbs{\Theta}_+ \mbs{\Psi}^\trans\right)
    {\left(\mbs{\Psi} \mbs{\Psi}^\trans\right)}^\dagger
    =
    \mbf{G}
    \mbf{H}^\dagger,
    \label{eq:edmdi-sol}
\end{equation}
where
\begin{equation}
    \mbf{G}
    =
    \frac{1}{q}
    \mbs{\Theta}_+ \mbs{\Psi}^\trans \in \mathbb{R}^{p_\vartheta \times p},
    \quad
    \mbf{H}
    =
    \frac{1}{q}
    \mbs{\Psi} \mbs{\Psi}^\trans \in \mathbb{R}^{p \times p}.
    \label{eq:edmdi-sol-scaled}
\end{equation}
Now, only a
${p \times p}$
pseudoinverse is required, rather than a
${p \times q}$
pseudoinverse.
To improve numerical conditioning, $\mbf{G}$ and $\mbf{H}$ are often scaled by
the number of snapshots $q$, as in~\cref{eq:edmdi-sol-scaled}.
Note that
${\mbf{H} = \mbf{H}^\trans > 0}$
if the columns of $\mbs{\Psi}$ are linearly independent.

\subsection{LMI reformulation of EDMD}
To incorporate regularizers and constraints in a modular fashion, the Koopman
operator regression problem is reformulated as a convex optimization problem
with LMI constraints. Recall that the Koopman matrix $\mbf{U}$
minimizes~\cref{eq:koopman-cost}. It therefore also minimizes
\begin{align}
    J(\mbf{U})
    &=
    \frac{1}{q}
    \|\mbs{\Theta}_+ - \mbf{U} \mbs{\Psi}\|_\frob^2%
    \label{eq:rescaled-koopman-cost}
    %
    =
    \frac{1}{q}
    \trace{\left(
        \left(\mbs{\Theta}_+ - \mbf{U} \mbs{\Psi}\right)
        {\left(\mbs{\Theta}_+ - \mbf{U} \mbs{\Psi}\right)}^\trans
    \right)}
    \\
    &=
    \trace{\left(%
        \frac{1}{q}
        \mbs{\Theta}_+ \mbs{\Theta}_+^\trans
        -
        \He{\mbf{U} \mbf{G}^\trans}
        +
        \mbf{U}
        \mbf{H}
        \mbf{U}^\trans
    \right)}
    \\
    &=
    c
    -
    2 \trace{\left(
        \mbf{U} \mbf{G}^\trans
    \right)}
    +
    \trace{\left(
        \mbf{U} \mbf{H} \mbf{U}^\trans
    \right)}, \label{eq:lmi-int-cost}
\end{align}
where
$c = \frac{1}{q} \mbs{\Theta}_+ \mbs{\Theta}_+^\trans$
is a scalar constant, 
$\mbf{G}$ and $\mbf{H}$ are defined in~\cref{eq:edmdi-sol-scaled},
and
${\He{\cdot} = {(\cdot)} + {(\cdot)}^\trans}$.
The minimization of~\cref{eq:lmi-int-cost} is equivalent to the minimization of
\begin{equation}
    J(\mbf{U}, \mbf{\nu}, \mbf{W})
    =
    c
    -
    2 \trace{\left(
        \mbf{U} \mbf{G}^\trans
    \right)}
    +
    \nu
\end{equation}
subject to
\begin{equation}
    \trace{(\mbf{W})} < \nu,
    \quad
    \mbf{W} > 0,
    \quad
    \mbf{U}\mbf{H}\mbf{U}^\trans < \mbf{W},\label{eq:pre-schur}
\end{equation}
where $\nu$ and $\mbf{W}$ are slack variables that allow the cost function to be
rewritten using LMIs~\cite[\S2.15.1]{caverly_2019_lmi}.
To rewrite the quadratic term in~\cref{eq:pre-schur} as an LMI, consider the
matrix decomposition ${\mbf{H} = \mbf{L} \mbf{L}^\trans}$.
The matrix $\mbf{L}$ can be found using a Cholesky factorization or
eigendecomposition of $\mbf{H}$, or a singular value decomposition of
$\mbs{\Psi}$.
Assuming the decomposition has been performed, the quadratic
term in the optimization problem becomes
\begin{equation}
    \mbf{W} - \mbf{U}\mbf{H}\mbf{U}^\trans
    =
    \mbf{W} - \mbf{U} \mbf{L} \mbf{L}^\trans \mbf{U}^\trans
    =
    \mbf{W} - \left(%
        \mbf{U} \mbf{L}
    \right)
    \mbf{1}
    {\left(%
        \mbf{U} \mbf{L}
    \right)}^\trans,\label{eq:pre-schur-2}
\end{equation}
where $\mbf{1}$ is the identity matrix.
Applying the Schur complement~\cite[\S2.3.1]{caverly_2019_lmi}
to~\cref{eq:pre-schur-2} yields
\begin{equation}
    \begin{bmatrix}
        \mbf{W} & \mbf{U} \mbf{L} \\
        \mbf{L}^\trans \mbf{U}^\trans & \mbf{1}
    \end{bmatrix}
    >
    0.
\end{equation}
The LMI form of the optimization problem is therefore
\begin{align}
    \min\;&
    J(\mbf{U}, \mbf{\nu}, \mbf{W})
    =
    c
    -2 \trace{\left(
        \mbf{U} \mbf{G}^\trans
    \right)}
    +
    \nu
    \\
    \mathrm{s.t.}\;&
    \trace{(\mbf{W})} < \nu,
    \quad
    \mbf{W} > 0,
    \quad
    \begin{bmatrix}
        \mbf{W} & \mbf{U} \mbf{L} \\
        \mbf{L}^\trans \mbf{U}^\trans & \mbf{1}
    \end{bmatrix}
    >
    0,
    \label{eq:noninverted-h}
\end{align}
where
${\mbf{H} = \mbf{L}^{}\mbf{L}^\trans}$.

As previously mentioned, the decomposition ${\mbf{H} = \mbf{L} \mbf{L}^\trans}$
can be realized via a Cholesky factorization, eigendecomposition, or singular
value decomposition. Using a Cholesky factorization directly gives $\mbf{L}$.
When using an eigendecomposition, ${\mbf{H} = \mbf{V} \mbs{\Lambda} \mbf{V}^\trans}$,
it follows that $\mbf{L} = \mbf{V}\sqrt{\mbs{\Lambda}}$.
Alternatively, using a singular value decomposition,
$\mbs{\Psi}
=
\mbf{Q}
\mbs{\Sigma}
\mbf{Z}^\trans$,
and substituting it into the definition of $\mbf{H}$
in~\cref{eq:edmdi-sol-scaled} yields
${\mbf{H} = \frac{1}{q} \mbf{Q}\mbs{\Sigma}^2\mbf{Q}^\trans}$.
It follows that
$\mbf{L} = \frac{1}{\sqrt{q}} \mbf{Q} \mbs{\Sigma}.$
A singular value decomposition is used to compute $\mbf{L}$ in the experiments
presented in this paper.

\section{Asymptotic Stability Constraint}\label{sec:srconst}

Since many systems of interest have asymptotically stable dynamics, it is
desirable to identify Koopman systems that share this property.
In~\cite{mamakoukas_2020_learning}, it is proven that an asymptotically stable
nonlinear system can only be represented accurately by an asymptotically stable
Koopman system, thus highlighting the importance of enforcing the asymptotic
stability property during the regression process.

However, in practice, it is possible to identify an unstable Koopman system from
measurements of an asymptotically stable system~\cite{mamakoukas_2020_learning}.
Even if the identified Koopman system is asymptotically stable in theory, the
eigenvalues of its $\mbf{A}$ matrix may be so close to the unit circle that it
is effectively unstable in practice.
One solution, presented in this section, is to constrain the largest eigenvalue
of $\mbf{A}$ to be strictly less than one in magnitude, within a desired
tolerance.

\subsection{Constraint formulation}
To ensure that the system defined by the Koopman matrix
${\mbf{U} = \begin{bmatrix} \mbf{A} & \mbf {B}\end{bmatrix}}$
is asymptotically stable, the eigenvalues of $\mbf{A}$ must be constrained
to have magnitude strictly less than one.
%
A modified Lyapunov constraint~\cite[\S1.4.4]{ghaoui_advances_2000}
\begin{align}
    \mbf{P} &> 0,
    \\
    \mbf{A}^\trans \mbf{P} \mbf{A} - \bar{\rho}^2\mbf{P} &< 0,%
    \label{eq:modified-lyap}
\end{align}
can be added to ensure that the magnitude of the largest eigenvalue of $\mbf{A}$
is no larger than ${0 < \bar{\rho} < 1}$.
Applying the Schur complement to~\cref{eq:modified-lyap} yields
\begin{align}
    \mbf{A}^\trans \mbf{P} \mbf{A} - \bar{\rho}^2\mbf{P} < 0
    &\iff
    \left(%
        \mbf{A}^\trans
        \mbf{P}
    \right)
    \mbf{P}^{-1}
    {\left(%
        \mbf{A}^\trans
        \mbf{P}
    \right)}^\trans
    -
    \bar{\rho}^2
    \mbf{P}
    <
    0
    \\
    &\iff
    -
    \bar{\rho}
    \mbf{P}
    -
    \left(%
        -\mbf{A}^\trans
        \mbf{P}
    \right)
    {\left(-\bar{\rho}\mbf{P}\right)}^{-1}
    {\left(%
        -\mbf{A}^\trans
        \mbf{P}
    \right)}^\trans
    <
    0
    \\
    &\iff
    \begin{bmatrix}
        - \bar{\rho}\,\mbf{P} & -\mbf{A}^\trans\mbf{P} \\
        -\mbf{P}^\trans\mbf{A} & - \bar{\rho}\,\mbf{P}
    \end{bmatrix}
    <
    0,\ -\bar{\rho}\,\mbf{P} < 0
    \\
    &\iff
    \begin{bmatrix}
        \bar{\rho}\,\mbf{P} & \mbf{A}^\trans\mbf{P} \\
        \mbf{P}^\trans\mbf{A} & \bar{\rho}\,\mbf{P}
    \end{bmatrix}
    >
    0,\ \mbf{P} > 0.\label{eq:lyap-bmi}
\end{align}
The full optimization problem with asymptotic stability constraint is therefore
\begin{align}
    \min\;&
    J(\mbf{U}, \mbf{\nu}, \mbf{W}, \mbf{P}; \bar{\rho})
    =
    c
    -2 \trace{\left(
        \mbf{U} \mbf{G}^\trans
    \right)}
    +
    \nu\label{eq:lyap-bmi-opt}
    \\
    \mathrm{s.t.}\;&
    \trace{(\mbf{W})} < \nu,
    \quad
    \mbf{W} > 0,
    \quad
    \begin{bmatrix}
        \mbf{W} & \mbf{U} \mbf{L} \\
        \mbf{L}^\trans \mbf{U}^\trans & \mbf{1}
    \end{bmatrix}
    >
    0,
    \quad
    \mbf{P} > 0,
    \quad
    \begin{bmatrix}
        \bar{\rho}\,\mbf{P} & \mbf{A}^\trans\mbf{P} \\
        \mbf{P}^\trans\mbf{A} & \bar{\rho}\,\mbf{P}
    \end{bmatrix}
    > 0,\label{eq:lyap-bmi-constr}
\end{align}
where
${\mbf{H} = \mbf{L} \mbf{L}^\trans}$
and
${\mbf{U} = \begin{bmatrix} \mbf{A} & \mbf {B}\end{bmatrix}}$.

Since both $\mbf{A}$ and $\mbf{P}$ are unknown, \cref{eq:lyap-bmi} includes a
BMI constraint. The optimization problem is therefore nonconvex and NP-hard. One
method to find a locally optimal solution is outlined
in~\cite{doroudchi_2018_decentralized}. First, assume $\mbf{P}$ is constant with
an initial guess of $\mbf{P} = \mbf{1}$ and solve
\crefrange{eq:lyap-bmi-opt}{eq:lyap-bmi-constr} as an LMI problem. Then, hold
the remaining variables constant using the solution from that optimization, and
solve a feasibility problem for $\mbf{P}$. Repeat this process until the cost
function stops changing significantly. Although this approach is rather simple,
it does result in an asymptotically stable Koopman system with reasonable
prediction error.

\subsection{Experimental results}

The effectiveness of the proposed asymptotic stability constraint is
demonstrated using experimental data collected from the Fatigue Structural
Testing Equipment Research (FASTER) platform at the National Research Council of
Canada (NRC)~\cite{fortune_system_2019}, which is used for aircraft fatigue
testing research.
\begin{figure}[htbp]
    \centering
    \begin{subfigure}[t]{0.5\textwidth}
        \centering
        \caption{}\label{fig:faster_eig}
        \includegraphics[width=\linewidth]{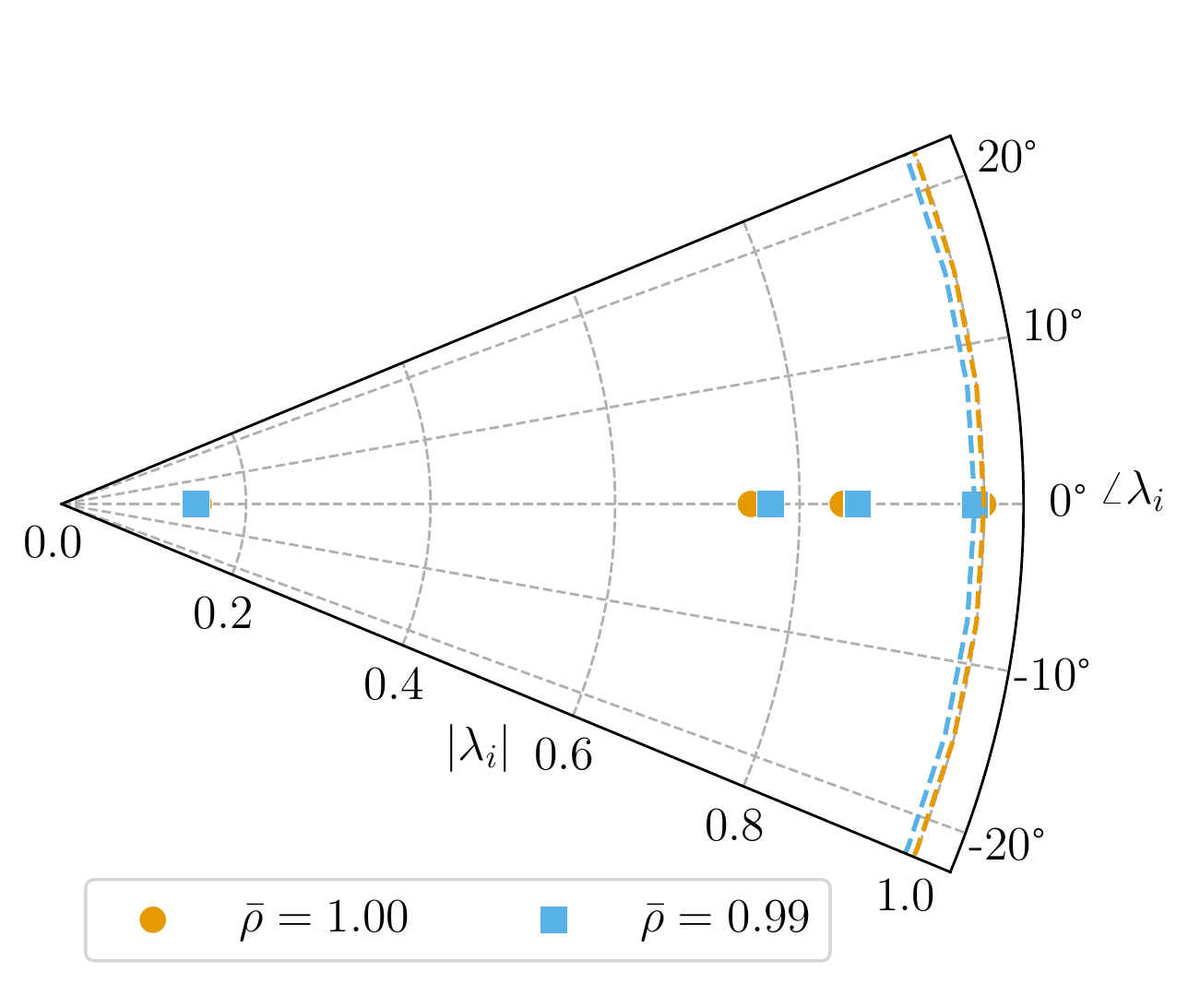}
    \end{subfigure}%
    \begin{subfigure}[t]{0.5\textwidth}
        \centering
        \caption{}\label{fig:faster_time}
        \includegraphics[width=\linewidth]{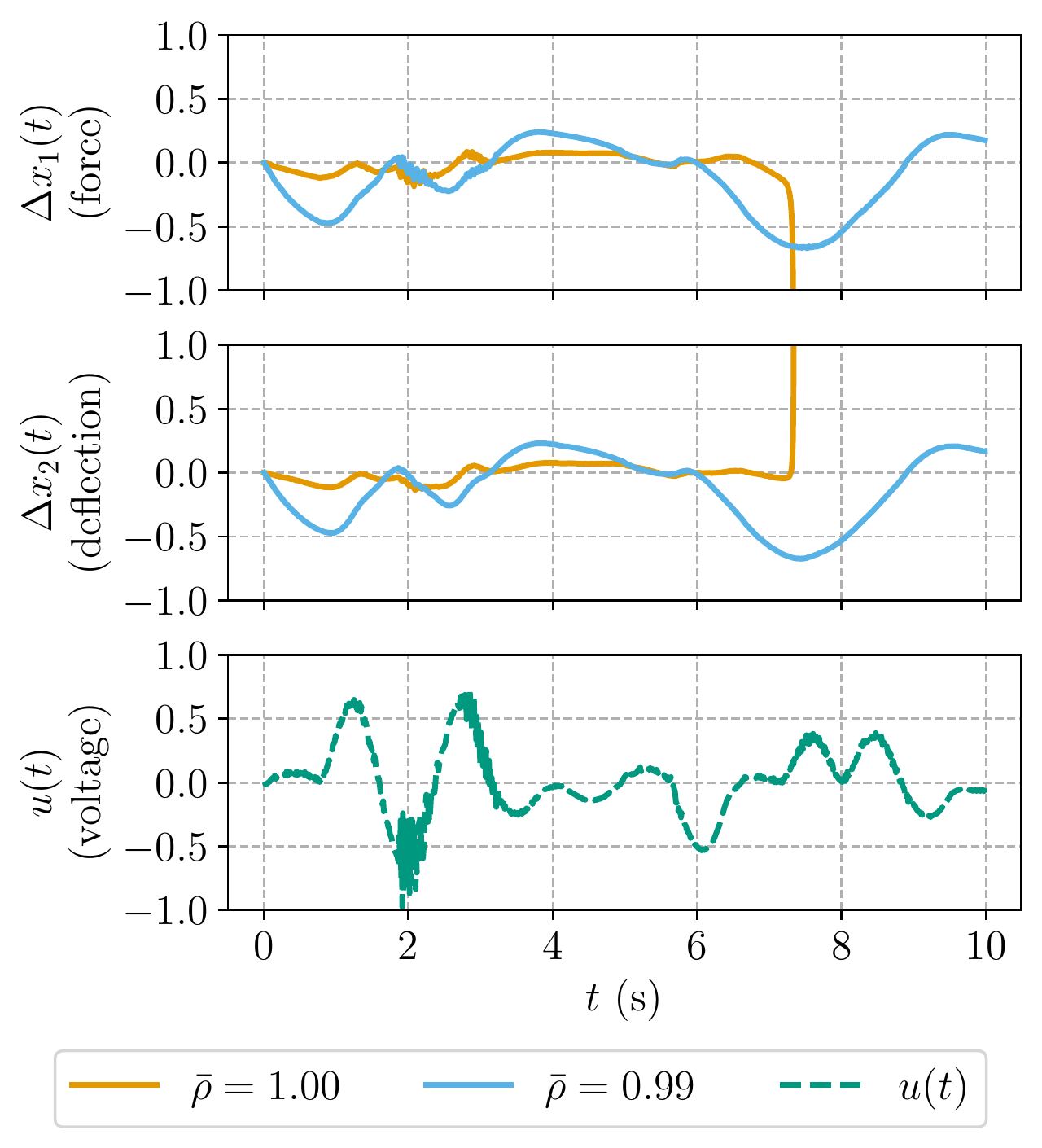}
    \end{subfigure}%
    \caption{(\subref{fig:faster_eig})~Eigenvalues and spectral radius
    constraints of Koopman $\mbf{A}$ matrices approximated from the FASTER
    dataset. The eigenvalues of $\mbf{A}$ satisfy their respective spectral
    radius constraints. Additionally, lowering the spectral radius constraint
    from ${\bar{\rho} = 1.00}$ to ${\bar{\rho} = 0.99}$ does not significantly
    alter the eigenvalues.
    (\subref{fig:faster_time})~Multi-step prediction error of Koopman systems
    approximated from the FASTER dataset. All units are normalized. States are
    recovered and re-lifted between prediction timesteps. Although both systems
    are asymptotically stable, only the system with ${\bar{\rho} = 0.99}$ is
    usable in practice, as the other system's response diverges due to the
    accumulation of numerical error.}\label{fig:faster}
\end{figure}
In the fatigue structural testing dataset used in this paper, the structure
under test is an aluminum-composite beam. During a test, the FASTER platform
applies a force to this beam using a hydraulic actuator, which is controlled by
a voltage. This input, the applied force, and the structure's deflection are
recorded at \SI{128}{\hertz}. The actuator voltage is determined by a linear
controller designed to track a reference force profile.
In this case, the dynamics of the controller are neglected and the actuator
voltage is considered to be exogenous.
All states and inputs in the
FASTER dataset are normalized. A photograph of this experimental setup can be
found in~\cite{fortune_system_2019}.

The Koopman lifting functions chosen for the FASTER platform are first- and
second-order monomials. The full lifting procedure consists of several steps to
improve numerical conditioning. First, all states and inputs are normalized so
that they do not grow when passed through the monomial lifting functions. Then,
all first- and second-order monomials of the state and input are computed.
Finally, the lifted states are standardized to ensure that they are evenly
weighted in the regression. To standardize the lifted states, their means
are subtracted and they are rescaled to have unit
variance~\cite[\S4.6.6]{james_introduction_2013}.
When using the identified Koopman matrices for prediction, the state and input
are lifted, then multiplied by the Koopman matrix. The state is then recovered
and re-lifted with the input for the next timestep. Using this prediction method
leads to the local error definition presented
in~\cite{mamakoukas_2020_learning}, which is used throughout this paper.

Given that the true FASTER system is asymptotically stable, it is crucial
to ensure that the identified Koopman system shares this property.
To demonstrate the impact of the asymptotic stability constraint, Koopman
matrices with maximum spectral radii of ${\bar{\rho} = 1.00}$ and ${\bar{\rho} =
0.99}$ are computed and compared to an unconstrained Koopman matrix computed
with standard EDMD.
\Cref{fig:faster_eig} shows the eigenvalues of the two constrained Koopman
$\mbf{A}$ matrices. In both cases, the eigenvalues indicate that the systems
obey their respective maximum spectral radii.
However, the multi-step prediction errors of the two Koopman systems in
\cref{fig:faster_time} show that only the system with the spectral radius
constraint of
${\bar{\rho} = 0.99}$
is usable in practice.
The other system produces prediction errors that diverge to infinity due to
the accumulation of numerical error.
The Koopman matrix without asymptotic stability constraint behaves identically
to the matrix with maximum spectral radius ${\bar{\rho} = 1.00}$, and is
therefore not shown \cref{fig:faster}.
%
%
By introducing a small amount of conservatism in the spectral radius constraint,
the identified Koopman system is rendered asymptotically stable in the face of
accumulated numerical error.
%

\section{System Norm Regularization}\label{sec:hinf}
While constraining the asymptotic stability of the identified Koopman system
helps ensure that the system's predictions are usable, it does not consider the
input-output properties of the system.
A system norm like the \Htwo{}~norm or the \Hinf{}~norm can be used to
regularize the system's gain, while also ensuring asymptotic stability for any
regularization coefficient.
Specifically, the existence of a finite \Htwo{}~or \Hinf{}~norm guarantees
the asymptotic stability of the resulting linear time-invariant (LTI)
system~\cite{zhou_robust_1995}.
Regularizing using the \Htwo{}~norm can be thought of as penalizing the average
system gain over all frequencies, while using the \Hinf{}~norm penalizes the
worst-case system gain.
As such, the use of the \Hinf{}~norm as a regularizer is explored next.
Weighted \Hinf{}~norms are also considered as regularizers, which allow the
regularization problem to be tuned in the frequency domain with weighting
functions.

\subsection{\texorpdfstring{$\mc{H}_\infty\!$}{H-infinity}~norm regularization}
Since approximating the Koopman matrix amounts to identifying a discrete-time
LTI system, it is natural to consider a system norm as a regularizer rather than
a matrix norm.
The Koopman representation of a nonlinear ODE can be
thought of as a discrete-time LTI system
${\mbs{\mc{G}}: \ell_{2e} \to \ell_{2e}}$,
where $\ell_{2e}$ is the extended inner product sequence
space~\cite[\S B.1.1]{green_linear_1994} and
${\mbf{U} = \begin{bmatrix}\mbf{A} & \mbf{B}\end{bmatrix}}$.
Consider the lifted output equation,
\begin{equation}
    \mbs{\zeta}_k
    =
    \mbf{C}
    \mbs{\vartheta}_k
    +
    \mbf{D}
    \mbs{\upsilon}_k,
\end{equation}
where $\mbf{C} \in \mathbb{R}^{p_\zeta \times p_\vartheta}$
and $\mbf{D} \in \mathbb{R}^{p_\zeta \times p_\upsilon}$.
In the simplest formulation,
${\mbf{C} = \mbf{1}}$,
and
${\mbf{D} = \mbf{0}}$.
The Koopman system is then
\begin{equation}
    \mbs{\mc{G}}
    \stackrel{\min}{\sim}
    \bma{c|c}
        \mbf{A} & \mbf{B} \\
        \hline 
        \mbf{C} & \mbf{D}
    \ema,
\end{equation}
where $\stackrel{\min}{\sim}$ denotes a minimal state-space
realization~\cite[\S3.2.1]{green_linear_1994}.
The Koopman system $\mbs{\mc{G}}$ has a corresponding discrete-time transfer
matrix~\cite[\S3.7]{zhou_robust_1995},
${\mbf{G}(z) = \mbf{C} (z\mbf{1} - \mbf{A})^{-1} \mbf{B} + \mbf{D}}$.

The $\mc{H}_\infty$ norm of $\mbs{\mc{G}}$ is the worst-case gain from
$\|\mbs{\upsilon}\|_2$ to $\|\mbs{\mc{G}}\mbs{\upsilon}\|_2$. That
is~\cite[\S B.1.1]{green_linear_1994}
\begin{equation}
    \|\mbs{\mc{G}}\|_\infty
    =
    \sup_{\mbs{\upsilon} \in \ell_{2}, \mbs{\upsilon} \neq \mbf{0}}
    \frac{\|\mbs{\mc{G}}\mbs{\upsilon}\|_2}{\|\mbs{\upsilon}\|_2},%
    \label{eq:hinf-norm}
\end{equation}
where $\ell_{2}$ is the inner product sequence
space~\cite[\S B.1.1]{green_linear_1994}. In the frequency domain, this
definition is equivalent to~\cite[\S B.1.1]{green_linear_1994}
\begin{equation}
    \|\mbs{\mc{G}}\|_\infty
    =
    \sup_{\theta \in (-\pi, \pi]}
    \bar{\sigma}\left(\mbf{G}(e^{j \theta})\right),%
    \label{eq:hinf-norm-f}
\end{equation}
where $\bar{\sigma}(\cdot)$ denotes the maximum singular value of a matrix.
In~\cref{eq:hinf-norm-f}, the transfer function is evaluated at
${z = e^{j \theta}}$, where
${\theta = 2 \pi \Delta t f}$
is the discrete-time frequency,
$\Delta t$ is the sampling timestep,
and $f$ is the continuous-time frequency.

With \Hinf{} norm regularization,
the cost function associated with the regression problem is
\begin{equation}
    J(\mbf{U}; \beta)
    =
    \|\mbs{\Theta}_+ - \mbf{U} \mbs{\Psi}\|_\frob^2
    +
    \beta \|\mbs{\mc{G}}\|_\infty,
    \label{eq:hinf-reg-cost}
\end{equation}
where $\beta$ is the regularization coefficient.
To integrate the \Hinf{}~norm into the regression problem, its LMI formulation
must be considered.
The inequality
${\|\mbs{\mc{G}}\|_\infty < \gamma}$
holds if and only if~\cite[\S3.2.2]{caverly_2019_lmi}
\begin{equation}
    \mbf{P} > 0,
    \quad
    \begin{bmatrix}
        \mbf{P} & \mbf{A} \mbf{P} & \mbf{B} & \mbf{0} \\
        \mbf{P}^\trans \mbf{A}^\trans & \mbf{P} & \mbf{0} &
            \mbf{P} \mbf{C}^\trans \\
        \mbf{B}^\trans & \mbf{0} & \gamma \mbf{1} & \mbf{D}^\trans \\
        \mbf{0} & \mbf{C} \mbf{P}^\trans & \mbf{D} & \gamma \mbf{1}
    \end{bmatrix}
    >
    0.
\end{equation}

The full optimization problem with \Hinf{} regularization is
\begin{align}
    \min\;&
    J(\mbf{U}, \mbf{\nu}, \mbf{W}, \gamma, \mbf{P}; \beta)
    =
    c
    -2 \trace{\left(
        \mbf{U} \mbf{G}^\trans
    \right)}
    +
    \nu
    +
    \frac{\beta}{q}\gamma\label{eq:hinf-norm-cost}
    \\
    \mathrm{s.t.}\;&
    \trace{(\mbf{W})} < \nu,
    \quad
    \mbf{W} > 0,
    \quad
    \begin{bmatrix}
        \mbf{W} & \mbf{U} \mbf{L} \\
        \mbf{L}^\trans \mbf{U}^\trans & \mbf{1}
    \end{bmatrix}
    >
    0,
    \\
    &\mbf{P} > 0,
    \quad
    \begin{bmatrix}
        \mbf{P} & \mbf{A} \mbf{P} & \mbf{B} & \mbf{0} \\
        \mbf{P}^\trans \mbf{A}^\trans & \mbf{P} & \mbf{0} &
            \mbf{P} \mbf{C}^\trans \\
        \mbf{B}^\trans & \mbf{0} & \gamma \mbf{1} & \mbf{D}^\trans \\
        \mbf{0} & \mbf{C} \mbf{P}^\trans & \mbf{D} & \gamma \mbf{1}
    \end{bmatrix}
    >
    0,%
    \label{eq:hinf-constraint}
\end{align}
where
${\mbf{H} = \mbf{L} \mbf{L}^\trans}$
and
${\mbf{U} = \begin{bmatrix} \mbf{A} & \mbf {B}\end{bmatrix}}$.

Like the asymptotic stability constraint proposed in \cref{sec:srconst},
\cref{eq:hinf-constraint} includes a BMI constraint in terms of the unknowns
$\mbf{P}$ and $\mbf{A}$. As such, the optimization problem in
\crefrange{eq:hinf-norm-cost}{eq:hinf-constraint} is nonconvex and NP-hard.
However, it can be solved using the same iterative procedure described in
\cref{sec:srconst}\cite{doroudchi_2018_decentralized}.

\subsection{Weighted \texorpdfstring{$\mc{H}_\infty\!$}{H-infinity}~norm regularization}
The \Hinf{}~norm used in~\cref{eq:hinf-norm-cost} can be weighted by cascading
$\mbs{\mc{G}}$ with another LTI system, $\mbs{\mc{G}}^\mathrm{w}$. For example,
choosing $\mbs{\mc{G}}^\mathrm{w}$ to be a highpass filter penalizes
system gains at high frequencies.
Weights can be cascaded before or after
$\mbs{\mc{G}}$ to weight the inputs, outputs, or both. Recall that
multi-input multi-output LTI systems do not commute.

Consider the weight
\begin{equation}
    \mbs{\mc{G}}^\mathrm{w}
    \stackrel{\min}{\sim}
    \bma{c|c}
        \mbf{A}^{\!\mathrm{w}} & \mbf{B}^\mathrm{w} \\
        \hline 
        \mbf{C}^\mathrm{w} & \mbf{D}^\mathrm{w}
    \ema,
\end{equation}
with state $\mbs{\vartheta}^\mathrm{w}$, input $\mbs{\upsilon}^\mathrm{w}$, and
output $\mbs{\zeta}^\mathrm{w}$.
Cascading $\mbs{\mc{G}}^\mathrm{w}$ after $\mbs{\mc{G}}$ yields the augmented
state-space system
\begin{align}
    \begin{bmatrix}
        \mbs{\vartheta}_{k+1} \\
        \mbs{\vartheta}^\mathrm{w}_{k+1}
    \end{bmatrix}
    &=
    \begin{bmatrix}
        \mbf{A} & \mbf{0} \\
        \mbf{B}^\mathrm{w} \mbf{C} & \mbf{A}^{\!\mathrm{w}}
    \end{bmatrix}
    \begin{bmatrix}
        \mbs{\vartheta}_k \\
        \mbs{\vartheta}^\mathrm{w}_k
    \end{bmatrix}
    +
    \begin{bmatrix}
        \mbf{B} \\
        \mbf{B}^\mathrm{w} \mbf{D}
    \end{bmatrix}
    \mbs{\upsilon}_k,\label{eq:weighted-ss-AB}
    \\
    \mbs{\zeta}^\mathrm{w}_k
    &=
    \begin{bmatrix}
        \mbf{D}^\mathrm{w} \mbf{C} & \mbf{C}^\mathrm{w}
    \end{bmatrix}
    \begin{bmatrix}
        \mbs{\vartheta}_k \\
        \mbs{\vartheta}^\mathrm{w}_k
    \end{bmatrix}
    +
    \mbf{D}^\mathrm{w} \mbf{D} \,
    \mbs{\upsilon}_k.
\end{align}
Minimizing the \Hinf{}~norm of the augmented system
\begin{equation}
    \mbs{\mc{G}}^\mathrm{w}\mbs{\mc{G}}
    \stackrel{\min}{\sim}
    \bma{cc|c}
        \mbf{A} &
        \mbf{0} &
        \mbf{B} \\
        \mbf{B}^\mathrm{w} \mbf{C} &
        \mbf{A}^{\!\mathrm{w}} &
        \mbf{B}^\mathrm{w} \mbf{D} \\
        \hline 
        \mbf{D}^\mathrm{w} \mbf{C} &
        \mbf{C}^\mathrm{w} &
        \mbf{D}^\mathrm{w} \mbf{D}
    \ema
\end{equation}
is equivalent to minimizing the weighted \Hinf{}~norm of the original system.
The choice of weighting function used can be viewed as another hyperparameter
in the regression problem.
%




Weighting the regression problem in the frequency domain comes at the cost of
increasing the dimension of the optimization problem. When cascading the weight
before $\mbs{\mc{G}}$, the dimension of $\mbs{\vartheta}^\mathrm{w}$ scales with
the dimension of $\mbs{\upsilon}$. When cascading after $\mbs{\mc{G}}$, the
dimension of $\mbs{\vartheta}^\mathrm{w}$ scales with the dimension of
$\mbs{\zeta}$. In the regression problems considered here, only post-weighting
is considered, since $\mbs{\zeta}$ has a much smaller dimension.

\subsection{Experimental results}
The unique advantages of the \Hinf{}~norm regularizer are demonstrated using the
soft robot dataset published alongside~\cite{bruder_data-driven_2021}
and~\cite{bruder_modeling_2019}. Unregularized EDMD and Tikhonov-regularized
EDMD~\cite{tikhonov_1995_numerical, dahdah_2021_linear}, two standard Koopman
matrix approximation methods, are compared with the asymptotic stability
constraint from \cref{sec:srconst} and the \Hinf{}~norm regularizer presented in
this section. The Koopman systems identified using these regression methods are
analyzed in terms of their prediction errors, system properties, and numerical
conditioning.
%

\begin{figure}[htbp]
    \centering
    \includegraphics[width=\linewidth]{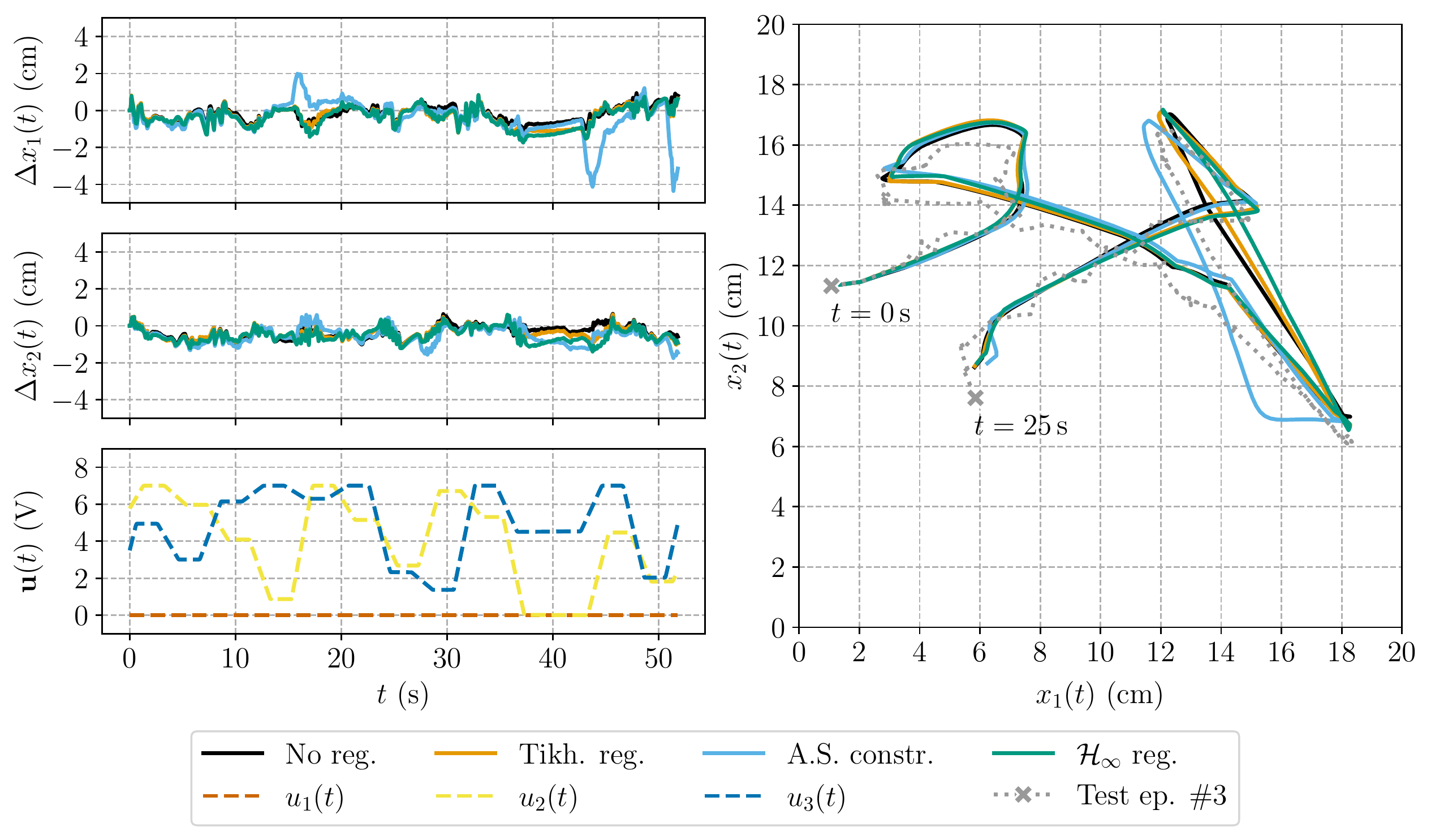}
    \caption{Multi-step prediction error and trajectory plot of the
    third test episode for Koopman systems approximated from the soft
    robot dataset. States are recovered and re-lifted between prediction
    timesteps. All Koopman systems have comparable prediction errors,
    with the exception of two large error spikes in the system with the
    asymptotic stability constraint.}\label{fig:soft_robot_error_traj}
\end{figure}
\begin{figure}[htbp]
    \centering
    \includegraphics[width=0.9\linewidth]{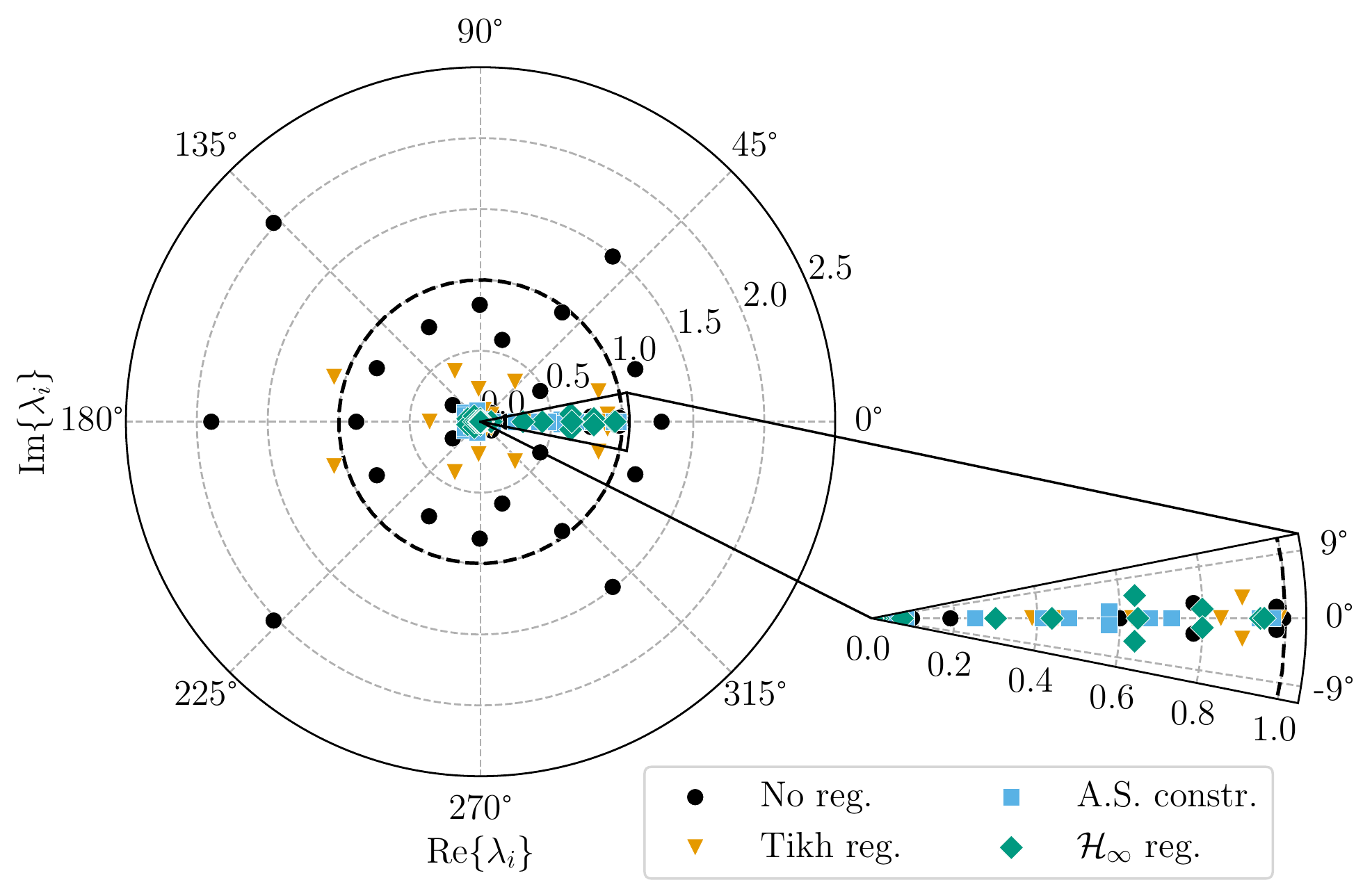}
    \caption{Eigenvalues of Koopman $\mbf{A}$ matrices approximated from the
    soft robot dataset. EDMD without regularization and EDMD with Tikhonov
    regularization both identify unstable systems, while the asymptotic
    stability constraint and \Hinf{}~norm regularizer yield asymptotically
    stable Koopman systems.}\label{fig:soft_robot_eig}
\end{figure}
\begin{figure}[htbp]
    \centering
    \includegraphics[width=\linewidth]{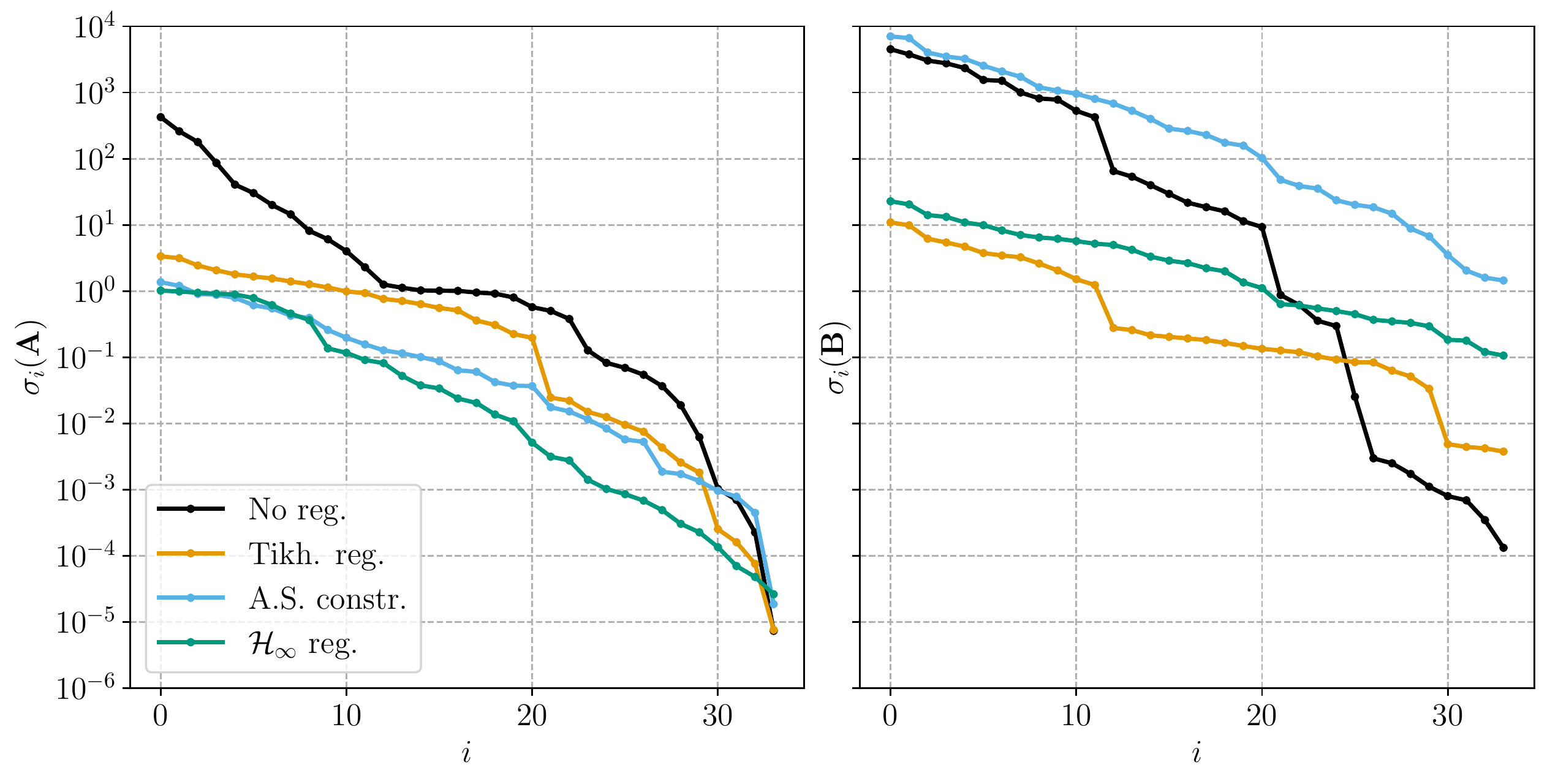}
    \caption{Singular values of Koopman $\mbf{A}$ and $\mbf{B}$ matrices
    approximated from the soft robot dataset plotted on a logarithmic scale. The
    $\mbf{A}$ and $\mbf{B}$ matrices computed using unregularized EDMD both have
    large singular values. EDMD with Tikhonov regularization decreases the
    singular values of both matrices, but does not yield an asymptotically
    stable system. With an asymptotic stability constraint, the singular values
    of $\mbf{A}$ are reduced, but the singular values of $\mbf{B}$ are
    unaffected and remain large. Using an \Hinf{}~norm regularizer reduces the
    singular values of both $\mbf{A}$ and $\mbf{B}$ significantly, yielding a
    better-conditioned Koopman matrix.}
    \label{fig:soft_robot_svd}
\end{figure}

Identifying a Koopman representation of a soft robot arm is a particularly
interesting problem, as its dynamics are not easily modelled from first
principles.
The soft robot under consideration consists of two flexible segments with a
laser pointer mounted at the end. The laser pointer projects a dot onto a board
positioned below the robot. The two states of the system are the Cartesian
coordinates of the dot on the board, as measured by a camera. The soft robot arm
is actuated by three pressure regulators, each controlled by a voltage. The dot
position and control voltages are recorded at \SI{12}{\hertz}.
Thirteen training episodes and four test episodes were recorded in this
manner. The third test episode is shown
in \cref{fig:soft_robot_error_traj}.
Photographs of this experimental setup, along with additional details, can be
found in~\cite{bruder_data-driven_2021} and~\cite{bruder_modeling_2019}.

The lifting functions chosen for the soft robot system consist of a time delay
step, followed by a third-order monomial transformation.
Although time delays do not, strictly speaking, meet the definition of a lifting
function outlined in \cref{sec:background}, they are commonly used in the lifted
states of Koopman identification problems~\cite{korda_2018_linear,
bruder_modeling_2019}.
As with the FASTER
dataset in \cref{sec:srconst}, the states and inputs are first
normalized. Then, the states and inputs are augmented with their delayed
versions, where the delay period is one timestep. Next, all first-, second-, and
third-order monomials are computed. Finally, the lifted states are standardized.
Since the time delay step occurs before the monomial lifting step, cross-terms
including delayed and non-delayed states and inputs occur in the lifted state.

Unregularized EDMD and Tikhonov-regularized EDMD are used as baselines for
comparison with the proposed regression methods. Tikhonov regularization
improves the numerical conditioning of $\mbf{U}$ by penalizing its squared
Frobenius norm~\cite{tikhonov_1995_numerical, dahdah_2021_linear}.
The regularizers used in this section have coefficients of ${\beta =
\num{7.5e-3}}$, while the asymptotic stability constraint has a maximum spectral
radius of ${\bar{\rho} = 0.999}$.
\Cref{fig:soft_robot_error_traj} shows the multi-step prediction errors of
the four Koopman systems for the third test episode of the dataset.
The prediction errors are comparable for all four Koopman systems, aside from
two large error spikes produced by the system with the asymptotic stability
constraint.

\begin{figure}[htb]
    \centering
    \includegraphics[width=0.75\linewidth]{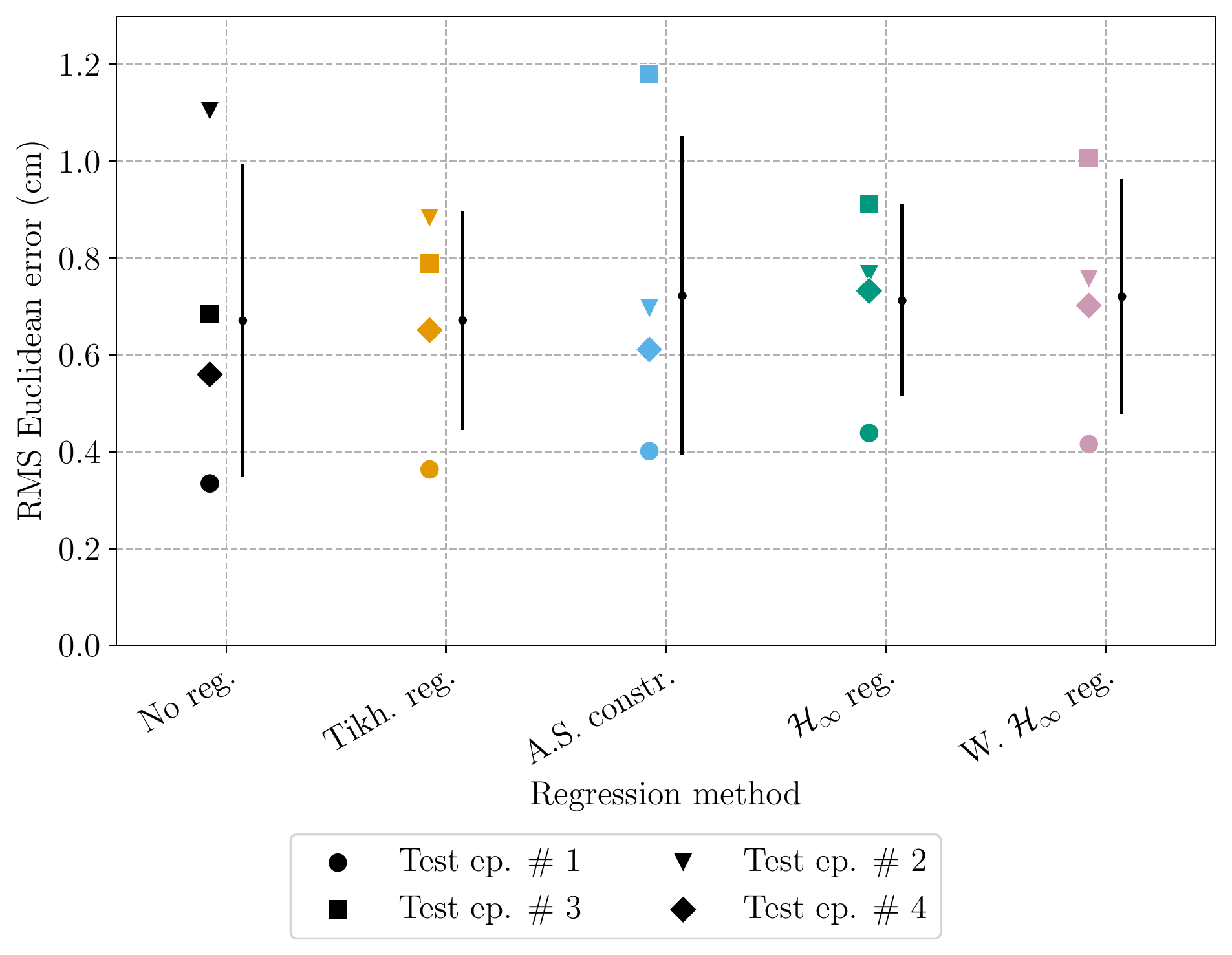}
    \caption{RMS Euclidean errors of Koopman systems approximated from the soft
    robot dataset. Error bars indicate mean and standard deviation of RMS error
    over the four test episodes. In terms of mean error, all identified
    Koopman systems perform similarly well. However, the system identified using
    \Hinf{}~norm regularizer performed more consistently throughout the
    test set.}\label{fig:soft_robot_scatter_by_method}
\end{figure}

One way to compare the resulting Koopman systems is to analyze the eigenvalues
of their $\mbf{A}$ matrices.
\Cref{fig:soft_robot_eig} shows that unregularized EDMD and Tikhonov-regularized
EDMD produce unstable Koopman systems, even though, as shown
in \cref{fig:soft_robot_scatter_by_method}, the multi-step prediction errors do
not happen to diverge in any test episodes.
As expected, EDMD with an \Hinf{}~norm regularizer and EDMD with an asymptotic
stability constraint both yield asymptotically stable Koopman systems.

\begin{figure}[htb]
    \centering
    \begin{subfigure}{0.5\textwidth}
        \centering
        \caption{}\label{fig:soft_robot_bode}
        \includegraphics[width=\linewidth]{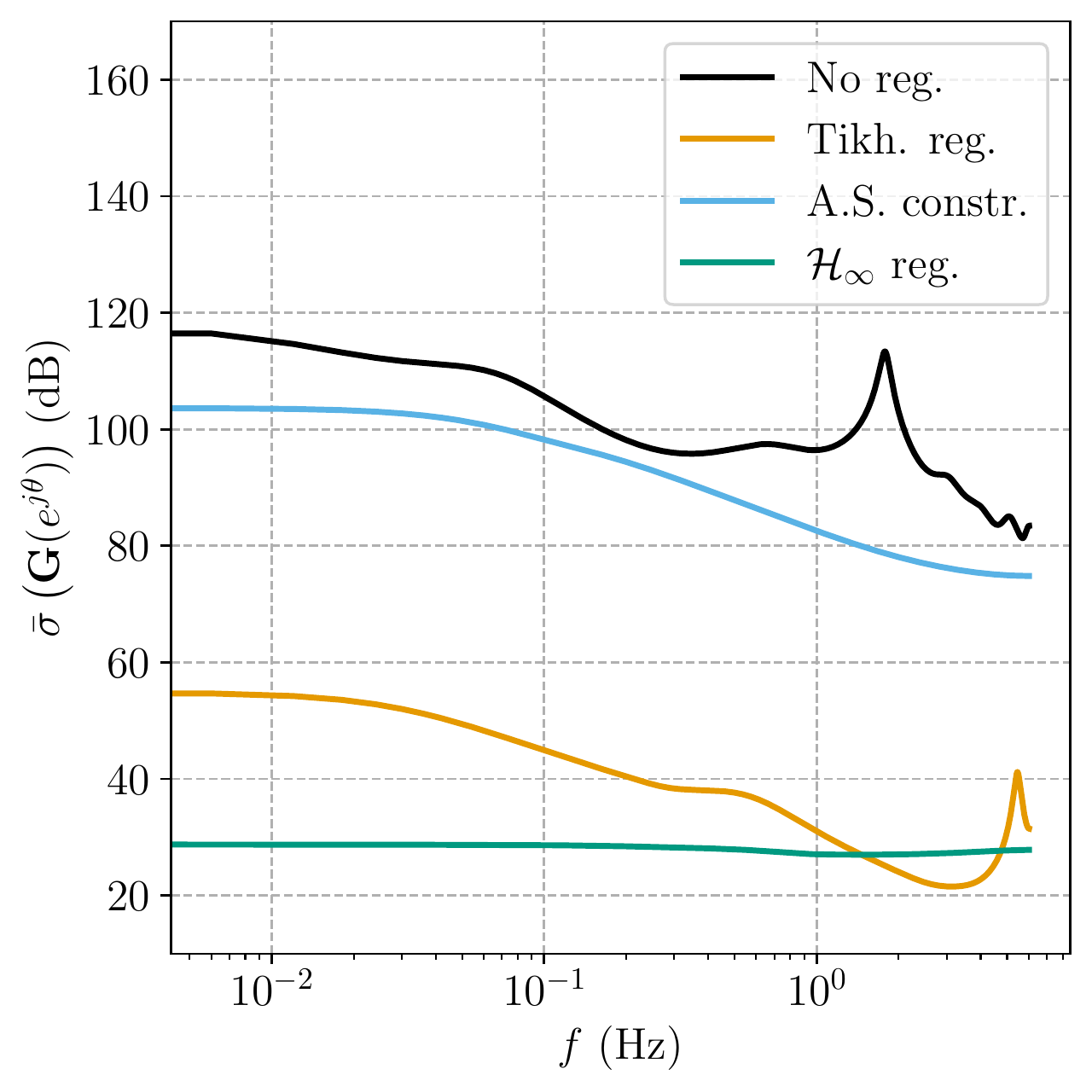}
    \end{subfigure}%
    \begin{subfigure}{0.5\textwidth}
        \centering
        \caption{}\label{fig:soft_robot_weights}
        \includegraphics[width=\linewidth]{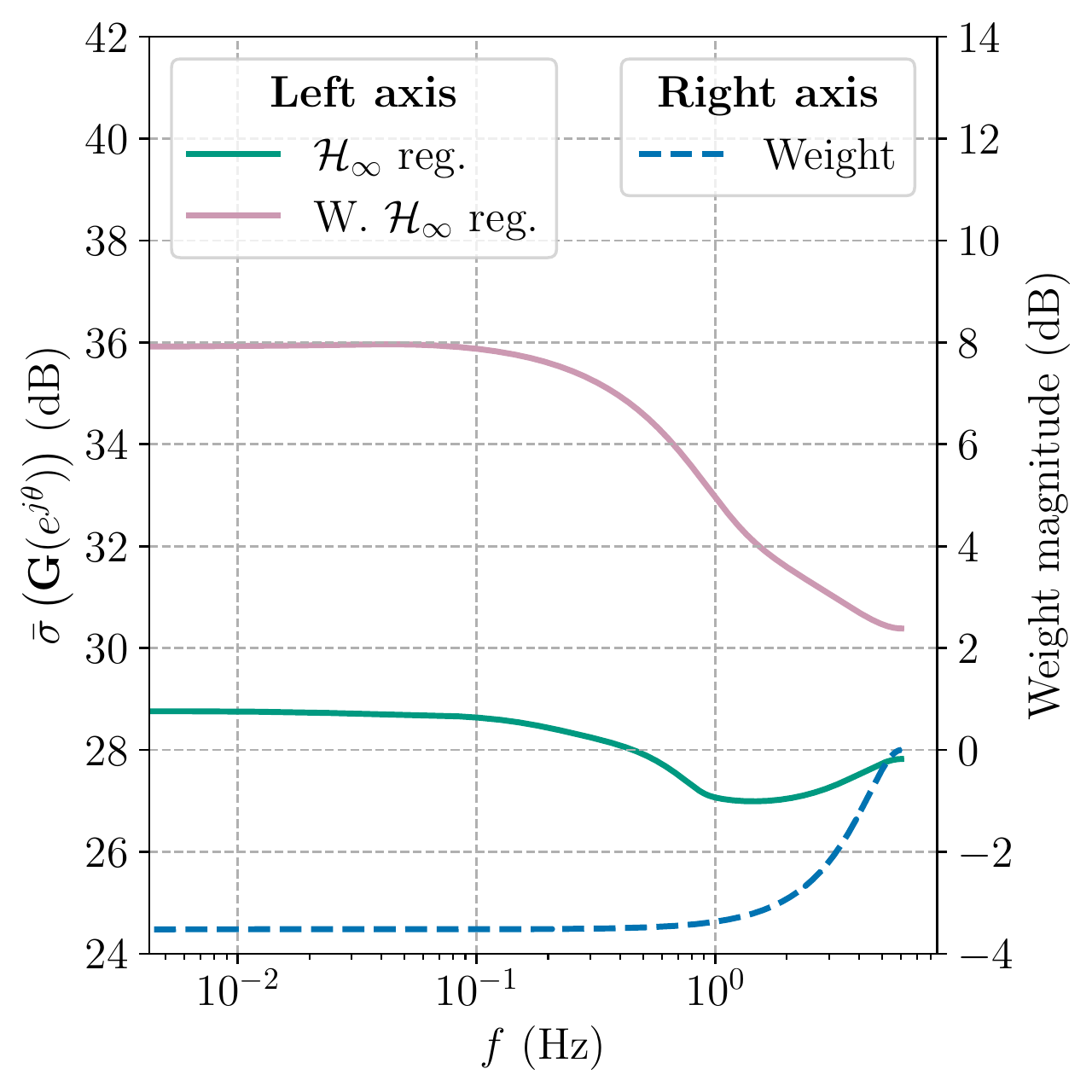}
    \end{subfigure}%
    \caption{(\subref{fig:soft_robot_bode})~Bode plot of Koopman systems
    approximated from the soft robot dataset.
    The Koopman system identified with unregularized EDMD has high gain, with a
    resonant peak at a high frequency.
    EDMD with Tikhonov regularization reduces the system's gain, but identifies
    an unstable system and retains an undesirable high-frequency resonant peak
    in the frequency response.
    Constraining the asymptotic stability of the system does not significantly
    reduce the gain compared to the unregularized system.
    Penalizing the \Hinf{}~norm of the Koopman system reduces its gain at all
    frequencies without compromising prediction error.
    (\subref{fig:soft_robot_weights})~Bode plot of unweighted and weighted
    Koopman systems approximated from the soft robot dataset, along with
    weighting function.
    The dashed line representing the weighting function uses the right axis,
    while the solid lines use the left axis.}\label{fig:soft_robot_freq}
\end{figure}

While the Koopman system identified with an asymptotic stability constraint is
indeed asymptotically stable, the system is not well-conditioned. To see this,
consider
\Cref{fig:soft_robot_svd}, which shows the singular values of the Koopman
$\mbf{A}$ and $\mbf{B}$ matrices. These singular values indicate the sizes of
the entries in each matrix.
With unregularized EDMD, both matrices have singular values on the order of
$10^3$. Using Tikhonov regularization decreases the singular values of both
$\mbf{A}$ and $\mbf{B}$, though it still yields an unstable system.
Constraining the spectral radius of $\mbf{A}$ greatly reduces the
singular values of $\mbf{A}$ but increases the singular values of $\mbf{B}$. The
numerical conditioning of this Koopman system is arguably worse, as the
$\mbf{A}$ and $\mbf{B}$ matrices contain entries of drastically different
scales.
Regularizing using the \Hinf{}~norm resolves this problem, as it reduces the
singular values in both matrices, yielding a better-conditioned, asymptotically
stable Koopman system with similar prediction error.
The key takeaway is that constraining the spectral radius of $\mbf{A}$ is not
sufficient to guarantee a well-conditioned Koopman matrix, as the constraint
does not directly impact $\mbf{B}$. Using the \Hinf{}~norm as a regularizer
considers the system as a whole, thus impacting both $\mbf{A}$ and
$\mbf{B}$, and reducing their entries to reasonable sizes.

Another way to compare the identified Koopman systems is by looking at their
frequency responses, which can be found by plotting the maximum singular value
of the transfer matrix at each frequency.
\Cref{fig:soft_robot_bode} shows the magnitude responses of the four Koopman
systems, and paints a similar picture to \cref{fig:soft_robot_svd}.
Unregularized EDMD yields a Koopman system with very high gain and a resonant
peak in the upper frequency range.
Incorporating Tikhonov regularization reduces the system's gain, but retains
the resonant peak at a high frequency.
Constraining the asymptotic stability of the system does not significantly
impact the system's gain, which, given the large singular values of $\mbf{B}$ in
\cref{fig:soft_robot_svd}, is not surprising.
However, regularizing using the \Hinf{}~norm directly penalizes the peak of the
Bode plot in \cref{fig:soft_robot_bode}, yielding a system with
significantly lower gain and similar prediction error.

The \Hinf{}~norm regularizer can be expanded upon by weighting the \Hinf{}~norm
using a highpass filter. This weighting function penalizes high gains at high
frequencies while allowing higher gains at low frequencies.
Penalizing gain at high frequencies is desirable because relevant system
dynamics typically occupy low frequencies, while high frequencies are
corrupted by measurement noise. Furthermore, causal physical systems have
frequency responses that roll off as frequency grows, since it is unrealistic
for a system to have infinite gain at infinitely high frequencies.
\Cref{fig:soft_robot_weights} demonstrates the impact of weighting the
\Hinf{}~norm regularizer with a highpass filter that has a zero at
\SI{4}{\hertz} and a pole slightly below \SI{6}{\hertz}. This weighted
regularizer yields a Koopman system with high gain at low frequencies and
decreasing gain at high frequencies.

Note that \cref{fig:soft_robot_freq} shows the frequency response of the Koopman
system in the lifted space, which is not the same as the ``frequency response
of the nonlinear system.'' Since the ultimate goal is to design linear
controllers in the lifted space, only the frequency response of the Koopman
system in the lifted space and the corresponding \Hinf{}~norm are relevant.

In \cref{fig:soft_robot_scatter_by_method}, the multi-step prediction errors of
the five identified Koopman systems are compared across each episode in the
test set.
Given that the laser pointer dot projected by the soft robot moves within a
circle of radius \SI{10}{\centi\metre}, it's practically meaningless to
distinguish each method based on mean prediction error alone.
However, the distribution of each identified system's prediction errors across
the test set clearly highlights the importance of regularization.
The Tikhonov regularizer, \Hinf{}~norm regularizer, and weighted \Hinf{}~norm
regularizer lead to systems with smaller standard deviations in the RMS error
over the test set. In contrast, EDMD without regularization and EDMD with
an asymptotic stability constraint lead to systems that perform inconsistently
over the test set. Comparing standard deviations indicates that the
regularized systems generalize better to previously unseen data, with the
\Hinf{}~norm regularizer performing most consistently over the four test
episodes.

The results in this section highlight the desirable properties of the proposed
\Hinf{}~norm regularizers. While all systems have comparable RMS prediction
errors on the test set, unregularized EDMD results in an unstable, poorly
conditioned Koopman system with large gain. While Tikhonov regularization
improves numerical conditioning, the resulting system is still unstable.
Conversely, constraining the asymptotic stability of the system does not improve
it numerical properties. Only the \Hinf{}~norm regularizers guarantee asymptotic
stability while improving the numerical conditioning of the system. These key
results are summarized in \cref{tab:cond}, which also highlights the difference
between the unweighted and weighted \Hinf{}~norm regularizers. Weighting the
\Hinf{}~norm regularizer further improves the condition numbers of $\mbf{A}$
and $\mbf{B}$.

\begin{table}[htbp]
    \caption{Comparison of regression methods through the condition numbers of
    their Koopman matrices and asymptotic stability guarantees.
    Only the \Hinf{}~regularizers guarantee asymptotic stability while
    significantly improving the condition number of the Koopman matrices. In
    this case, weighting the \Hinf{}~norm further improves
    $\mathrm{cond}(\mbf{A})$ and $\mathrm{cond}(\mbf{B})$.}\label{tab:cond}
    \centering
    \setlength{\tabcolsep}{12pt}
    \def\arraystretch{1.1}
    \begin{tabularx}{\linewidth}{llll}
        \toprule
        \tabhead{regression method} & \tabhead{$\mathrm{cond}(\mbf{A})$} & \tabhead{$\mathrm{cond}(\mbf{B})$} & \tabhead{asymptotic stability} \\
        no regularization & $5.77\times10^7$ & $3.40\times10^7$ & no \\
        \midrule
        Tikhonov regularization & $4.39\times10^5$ & $2.90\times10^3$ & no \\
        \midrule
        asymptotic stability constraint & $7.32\times10^4$ & $4.87\times10^3$ & yes \\
        \midrule
        $\mathcal{H}_\infty$ regularization & $3.87\times10^4$ & $2.14\times10^2$ & yes \\
        \midrule
        weighted $\mathcal{H}_\infty$ regularization & $1.69\times10^3$ & $5.43\times10^1$ & yes \\
        \botrule
    \end{tabularx}
\end{table}

\section{Reducing the Size of the Regression Problem}\label{sec:dmdc}
As the number of lifting functions required grows, so does the size of the
optimization problem. With several hundred lifting functions, finding a solution
can take days and consume an intractable amount of memory. To combat this
limitation, an approach reminiscent of DMDc is now presented.

\subsection{DMD with control}
DMD with control~\cite{proctor_2014_dynamic}
reduces the dimension of the Koopman matrix regression problem when the dataset
contains many more lifted states than time snapshots
(\ie{}, ${p \gg q}$)~\cite[\S10.3]{kutz_dynamic_2016}.
In DMDc, the Koopman matrix is projected onto the left singular vectors of
$\mbs{\Theta}_+$. The size of the problem is then controlled by retaining only
the $\hat{r}$ largest singular values in the SVD.
Consider the truncated singular value decomposition
${\mbs{\Theta}_+ \approx \mbfhat{Q} \mbshat{\Sigma} \mbfhat{Z}^\trans}$,
where
${\mbfhat{Q} \in \mathbb{R}^{p_\vartheta \times \hat{r}}}$,
${\mbshat{\Sigma} \in \mathbb{R}^{\hat{r} \times \hat{r}}}$, and
${\mbfhat{Z} \in \mathbb{R}^{q \times \hat{r}}}$.
Instead of solving for
${\mbf{U} = \begin{bmatrix}\mbf{A} & \mbf{B}\end{bmatrix}}$,
the regression problem is written in terms of~\cite{proctor_2014_dynamic}
\begin{equation}
    \mbfhat{U}
    =
    \begin{bmatrix}
        \mbfhat{Q}^\trans\mbf{A}\mbfhat{Q} &
        \mbfhat{Q}^\trans\mbf{B}
    \end{bmatrix}
    =
    \mbfhat{Q}^\trans
    \mbf{U}
    \!
    \begin{bmatrix}
        \mbfhat{Q} & \mbf{0} \\
        \mbf{0} & \mbf{1}
    \end{bmatrix},
    \label{eq:u-hat-def}
\end{equation}
where
${\mbfhat{U} \in \mathbb{R}^{\hat{r} \times \hat{r} + p_\upsilon}}$
is significantly smaller than
${\mbf{U} \in \mathbb{R}^{p_\vartheta \times p}}$.
The least-squares solution to the Koopman
matrix~\cref{eq:koopman-soln} can be written as
\begin{equation}
    \mbf{U}
    =
    \mbs{\Theta}_+
    \mbftilde{Z}
    \mbstilde{\Sigma}^\dagger
    \mbftilde{Q}^\trans,%
    \label{eq:koopman-soln-svd}
\end{equation}
where
${\mbs{\Psi} \approx \mbftilde{Q} \mbstilde{\Sigma} \mbftilde{Z}^\trans}$.
The number of singular values retained in this SVD is denoted
$\tilde{r}$. Thus
${\mbftilde{Q} \in \mathbb{R}^{p \times \tilde{r}}}$,
${\mbstilde{\Sigma} \in \mathbb{R}^{\tilde{r} \times \tilde{r}}}$, and
${\mbftilde{Z} \in \mathbb{R}^{q \times \tilde{r}}}$.
The standard solution to the DMDc problem is obtained by
substituting~\cref{eq:koopman-soln-svd}
into~\cref{eq:u-hat-def}, yielding
\begin{equation}
    \mbfhat{U}
    =
    \mbfhat{Q}^\trans
    \mbs{\Theta}_+
    \mbftilde{Z}
    \mbstilde{\Sigma}^\dagger
    \mbftilde{Q}^\trans
    %
    \begin{bmatrix}
        \mbfhat{Q} & \mbf{0} \\
        \mbf{0} & \mbf{1}
    \end{bmatrix}.
\end{equation}
The SVD dimensions $\hat{r}$ and $\tilde{r}$ are a design choice. A common
approach is the hard-thresholding algorithm described
in~\cite{gavish_2014_optimal}. Note that typically
${\hat{r} < \tilde{r}}$~\cite[\S6.1.3]{kutz_dynamic_2016}.

\subsection{LMI reformulation of DMDc}
To reformulate DMDc as a convex optimization problem with LMI constraints,
the cost function is rewritten in terms of $\mbfhat{U}$ instead of
$\mbf{U}$.
Recall that the rescaled Koopman cost function~\cref{eq:rescaled-koopman-cost}
can be written as
\begin{equation}
    \min J(\mbf{U})
    =
    \frac{1}{q}
    \trace{\left(%
        \mbs{\Theta}_+^{} \mbs{\Theta}_+^\trans
        - \He{\mbf{U}\mbs{\Psi}\mbs{\Theta}_+^\trans}
        + \mbf{U}\mbs{\Psi}\mbs{\Psi}^\trans\mbf{U}^\trans
    \right)}.%
    \label{eq:rescaled-koopman-cost-rewritten}
\end{equation}
With the introduction of a slack
variable,~\cref{eq:rescaled-koopman-cost-rewritten}
becomes~\cite[\S2.15.1]{caverly_2019_lmi}
\begin{align}
    \min\;&
    J(\mbf{U}, \mbf{W}) = \frac{1}{q} \trace{(\mbf{W})}%
    \\
    \mathrm{s.t.}\;&
    \mbf{W} > 0,%
    \quad
    \mbs{\Theta}_+^{} \mbs{\Theta}_+^\trans
    - \He{\mbf{U}\mbs{\Psi}\mbs{\Theta}_+^\trans}
    + \mbf{U}\mbs{\Psi}\mbs{\Psi}^\trans\mbf{U}^\trans < \mbf{W}.%
\end{align}
Substituting the SVDs of $\mbs{\Theta}_+$ and $\mbs{\Psi}$ into the optimization
problem yields
\begin{align}
    \min\;&
    J(\mbf{U}, \mbf{W}) = \frac{1}{q} \trace{(\mbf{W})}%
    \label{eq:congruence-cost}
    \\
    \mathrm{s.t.}\;&
    \mbf{W} > 0,%
    \quad
    \mbfhat{Q}\mbshat{\Sigma}^2\mbfhat{Q}^\trans
    - \He{%
        \mbf{U}\mbftilde{Q}\mbstilde{\Sigma}\mbftilde{Z}^\trans%
        \mbfhat{Z}\mbshat{\Sigma}\mbfhat{Q}^\trans%
    }
    + \mbf{U}\mbftilde{Q}\mbstilde{\Sigma}^2\mbftilde{Q}^\trans\mbf{U}^\trans
    < \mbf{W}.%
    \label{eq:congruence-2}
\end{align}
Next, consider the projection
of~\cref{eq:congruence-cost} and~\cref{eq:congruence-2}
onto the column space of $\mbfhat{Q}$, denoted $\mc{R}(\mbfhat{Q})$.
Recall that $\mbf{W} > 0$ is equivalent to
\begin{equation}
    \mbs{\vartheta}^\trans
    \mbf{W}
    \mbs{\vartheta} > 0,\ \forall
    \mbs{\vartheta} \neq \mbf{0} \in \mathbb{R}^{p_\vartheta \times 1}.%
    \label{eq:alternate-lmi-form}
\end{equation}
Since~\cref{eq:alternate-lmi-form} holds over all of
$\mathbb{R}^{p_\vartheta \times 1}$,
it must also hold over the subspace
$\mc{R}(\mbfhat{Q})$. Let the vectors in $\mc{R}(\mbfhat{Q})$ be parameterized
by
\begin{equation}
    \mbs{\vartheta} = \mbfhat{Q} \mbshat{\vartheta},%
    \label{eq:def-theta-hat}
\end{equation}
where
${\mbshat{\vartheta} \in \mathbb{R}^{\hat{r} \times 1}}$.
Substituting~\cref{eq:def-theta-hat} into~\cref{eq:alternate-lmi-form} yields
\begin{equation}
    \mbshat{\vartheta}^\trans
    \mbfhat{Q}^\trans
    \mbf{W}
    \mbfhat{Q}
    \mbshat{\vartheta} > 0,\ \forall
    \mbshat{\vartheta} \neq \mbf{0} \in \mathbb{R}^{\hat{r} \times 1},%
    \label{eq:alternate-lmi-form-hat}
\end{equation}
which is equivalent to
\begin{equation}
    \mbfhat{W} = \mbfhat{Q}^\trans\mbf{W}\mbfhat{Q} > 0
\end{equation}
over $\mc{R}(\mbfhat{Q})$. Applying the same logic to~\cref{eq:congruence-2}
yields
\begin{equation}
    \mbshat{\Sigma}^2
    - \He{%
        \mbfhat{Q}^\trans\mbf{U}\mbftilde{Q}%
        \mbstilde{\Sigma}\mbftilde{Z}^\trans%
        \mbfhat{Z}\mbshat{\Sigma}%
    }
    + \mbfhat{Q}^\trans\mbf{U}\mbftilde{Q}%
    \mbstilde{\Sigma}^2%
    \mbftilde{Q}^\trans\mbf{U}^\trans\mbfhat{Q}
    < \mbfhat{W},
\end{equation}
where the fact that
${\mbfhat{Q}^\trans \mbfhat{Q} = \mbf{1}}$
has been used.

To further simplify the problem, it is advantageous to
rewrite~\cref{eq:congruence-cost} in terms of $\mbfhat{W}$. To accomplish this,
first recall that the trace of a matrix is equal to the sum of its eigenvalues.
The eigenvalue problem for $\mbf{W}$ is
\begin{equation}
    \mbf{W} \mbf{v}_i = \lambda_i \mbf{v}_i.%
    \label{eq:eigendecomp}
\end{equation}
Projecting~\cref{eq:eigendecomp} onto $\mc{R}(\mbfhat{Q})$
by substituting
${\mbf{v}_i = \mbfhat{Q} \mbfhat{v}_i}$,
then premultiplying the result by $\mbfhat{Q}^\trans$, yields
\begin{equation}
    \mbfhat{Q}^\trans
    \mbf{W} \mbfhat{Q} \mbfhat{v}_i
    =
    \lambda_i \mbfhat{v}_i.
\end{equation}
Thus, $\mbf{W}$ and $\mbfhat{W}$ share the same eigenvalues for
eigenvectors in $\mc{R}(\mbfhat{Q})$, which indicates that minimizing
$\trace(\mbfhat{W})$ is equivalent to minimizing $\trace(\mbf{W})$
in that subspace.

The full regression problem projected onto $\mc{R}(\mbfhat{Q})$ is therefore
\begin{align}
    \min\;&
    J(\mbf{U}, \mbfhat{W}) = \frac{1}{q} \trace{(\mbfhat{W})}
    \\
    \mathrm{s.t.}\;&
    \mbfhat{W} > 0,
    \quad
    \mbshat{\Sigma}^2
    - \He{%
        \mbfhat{Q}^\trans\mbf{U}\mbftilde{Q}%
        \mbstilde{\Sigma}\mbftilde{Z}^\trans%
        \mbfhat{Z}\mbshat{\Sigma}%
    }
    + \mbfhat{Q}^\trans\mbf{U}\mbftilde{Q}%
    \mbstilde{\Sigma}^2%
    \mbftilde{Q}^\trans\mbf{U}^\trans\mbfhat{Q}
    < \mbfhat{W}.
\end{align}
Substituting $\mbfhat{U}$ from~\cref{eq:u-hat-def} into the optimization
problem yields
\begin{align}
    \min\;&
    J(\mbfhat{U}, \mbfhat{W}) = \frac{1}{q} \trace{(\mbfhat{W})}
    \\
    \mathrm{s.t.}\;&
    \mbfhat{W} > 0,
    \quad
    \mbshat{\Sigma}^2
    - \He{%
        \mbfhat{U}\mbfbar{Q}%
        \mbstilde{\Sigma}\mbftilde{Z}^\trans%
        \mbfhat{Z}\mbshat{\Sigma}%
    }
    + \mbfhat{U}\mbfbar{Q}%
    \mbstilde{\Sigma}^2%
    \mbfbar{Q}^\trans\mbfhat{U}^\trans
    < \mbfhat{W},%
    \label{eq:pre-schur-3}
\end{align}
where
\begin{equation}
    \mbfbar{Q}
    =
    \begin{bmatrix}
        \mbfhat{Q} & \mbf{0} \\
        \mbf{0} & \mbf{1}
    \end{bmatrix}^\trans
    \mbftilde{Q}.
\end{equation}
Applying the Schur complement to~\cref{eq:pre-schur-3} yields the LMI
formulation of DMDc,
\begin{align}
    \min\;&
    J(\mbfhat{U}, \mbfhat{W}) = \frac{1}{q} \trace{(\mbfhat{W})}
    \\
    \mathrm{s.t.}\;&
    \mbfhat{W} > 0,
    \quad
    \begin{bmatrix}
        -\mbfhat{W}
        + \mbshat{\Sigma}^2
        - \He{%
            \mbfhat{U}\mbfbar{Q}%
            \mbstilde{\Sigma}\mbftilde{Z}^\trans%
            \mbfhat{Z}\mbshat{\Sigma}%
        }
        &
        \mbfhat{U}\mbfbar{Q} \mbstilde{\Sigma}
        \\
        \mbstilde{\Sigma} \mbfbar{Q}^\trans\mbfhat{U}^\trans
        &
        -\mbf{1}
    \end{bmatrix} < 0.
\end{align}
This is now a significantly smaller optimization problem, as its size is
controlled by the truncation of the SVD of $\mbs{\Theta}_+$.
Reducing the first dimension of $\mbfhat{U}$ also reduces the dimension of
the slack variable $\mbfhat{W}$.

\subsection{Constraints and regularization}
The projection in~\cref{eq:def-theta-hat} defines a new Koopman system,
\begin{align}
    \mbshat{\vartheta}_{k+1}
    &=
    \mbfhat{A} \mbshat{\vartheta}_{k}
    + \mbfhat{B} \mbs{\upsilon}_k,%
    \label{eq:new-state}
    \\
    \mbs{\zeta}_k
    &=
    \mbfhat{C}
    \mbshat{\vartheta}_k
    +
    \mbf{D}
    \mbs{\upsilon}_k,%
    \label{eq:new-output}
\end{align}
where $\mbfhat{C} = \mbf{C} \mbfhat{Q}$.
The asymptotic stability constraint discussed in \cref{sec:srconst} and
the \Hinf{}~norm regularizers discussed in \cref{sec:hinf} can be equally
applied to the projected system in~\cref{eq:new-state}
and~\cref{eq:new-output} to ensure that the projected system of smaller
dimension has the desired stability and frequency response characteristics.

\subsection{Experimental results}

The properties of the asymptotic stability constraint from
\cref{sec:srconst} and the \Hinf{}~norm regularizer from
\cref{sec:hinf} are now compared when applied to the EDMD and DMDc
regression problems. The same dataset and experimental setup as
\cref{sec:hinf} is used here.

\begin{figure}[htbp]
    \centering
    \includegraphics[width=\linewidth]{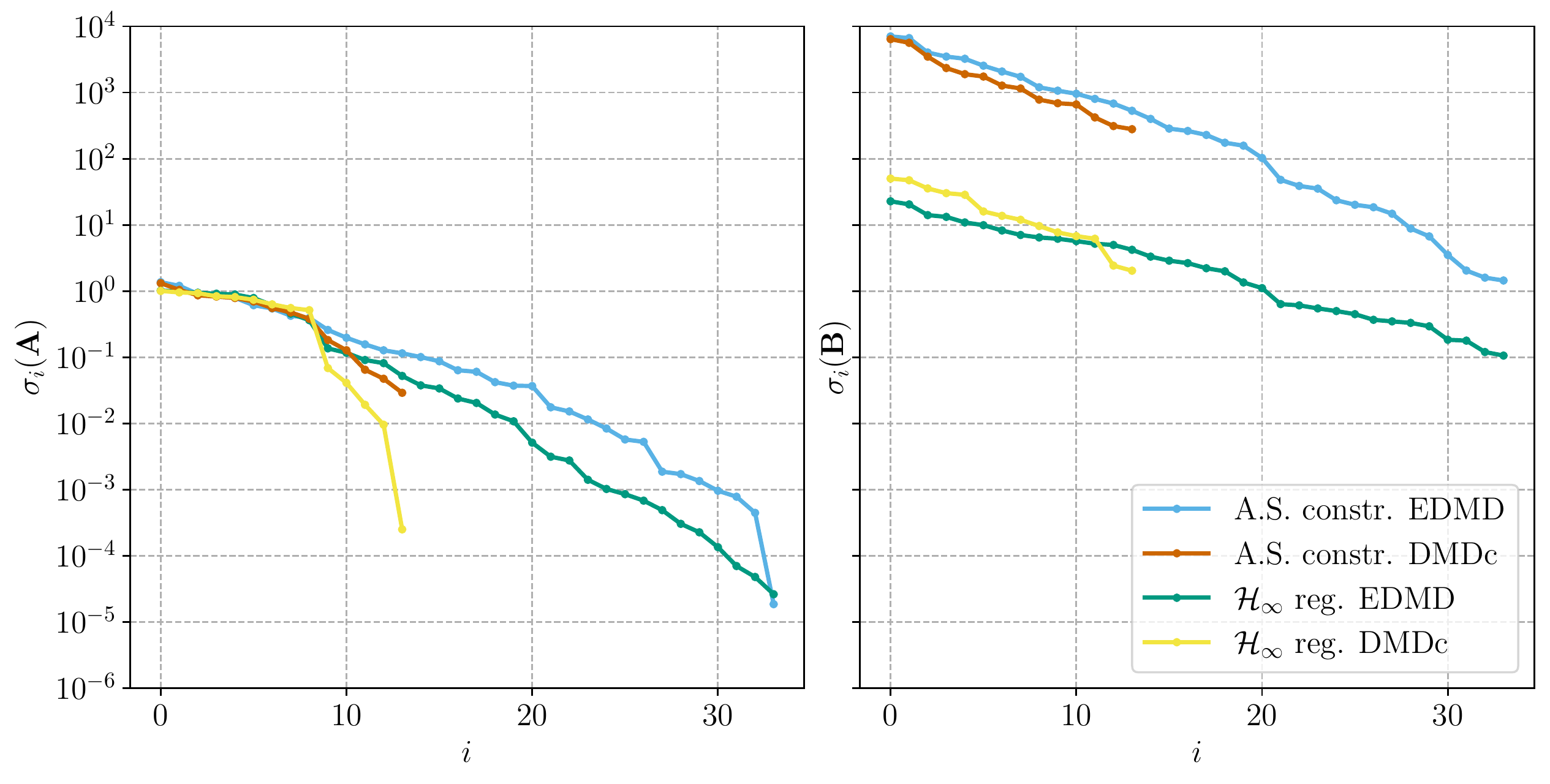}
    \caption{Singular values of Koopman $\mbf{A}$ and $\mbf{B}$ matrices
    approximated from the soft robot dataset using EDMD and DMDc regressors.
    Singular values smaller than $10^{-12}$ are not shown.
    Note the logarithmic scale.
    While the EDMD methods retain all 34 singular values, the DMDc methods
    truncate all but the first 14.
    The singular values retained by the DMDc methods are close to the
    corresponding singular values computed by the EDMD methods.
    }
    \label{fig:soft_robot_dmdc_svd}
\end{figure}
\begin{figure}[htbp]
    \centering
    \begin{subfigure}[t]{0.5\textwidth}
        \centering
        \caption{}\label{fig:soft_robot_dmdc_bode}
        \includegraphics[width=\linewidth]{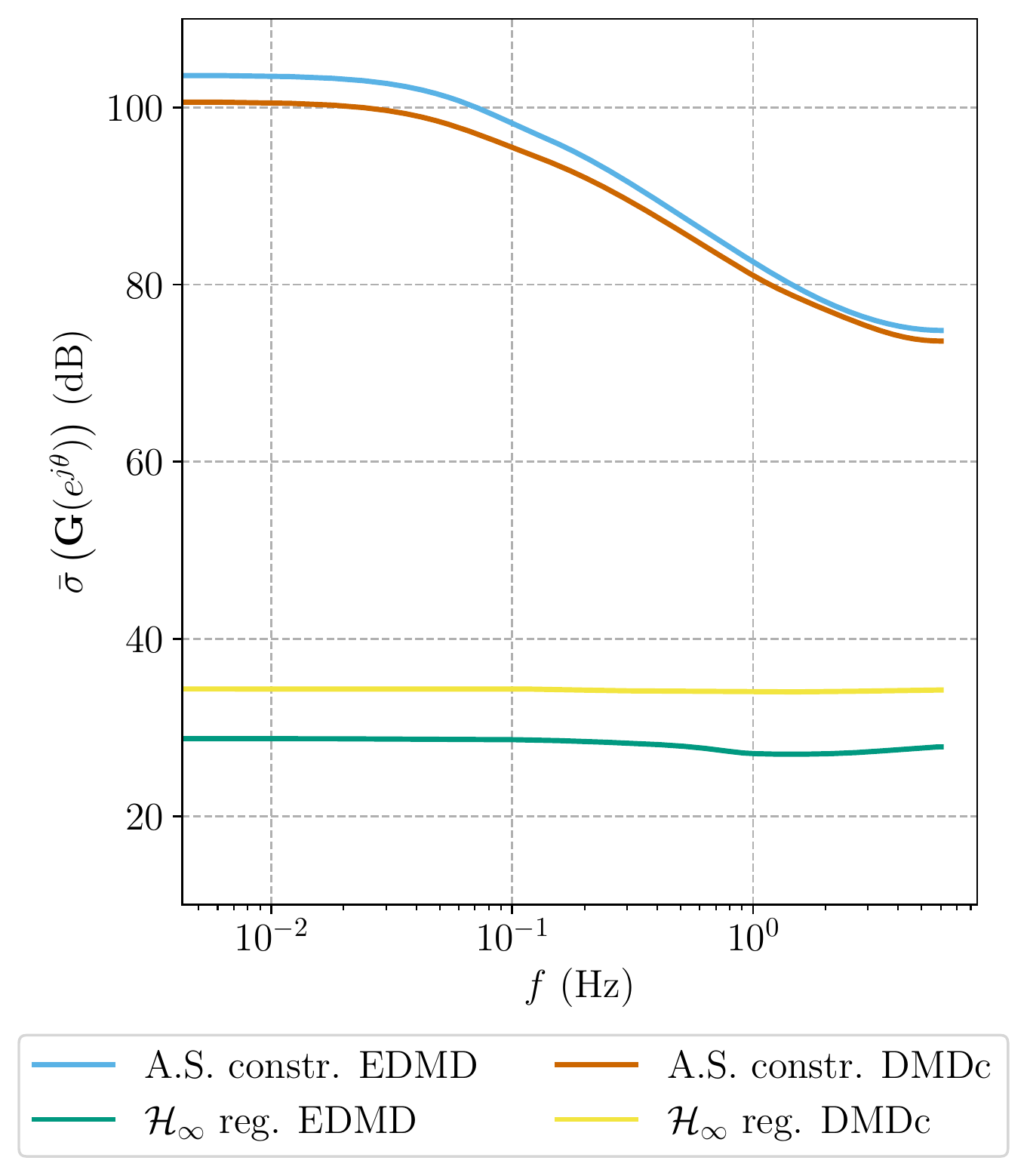}
    \end{subfigure}%
    \begin{subfigure}[t]{0.5\textwidth}
        \centering
        \caption{}\label{fig:soft_robot_scatter_dmdc}
        \includegraphics[width=\linewidth]{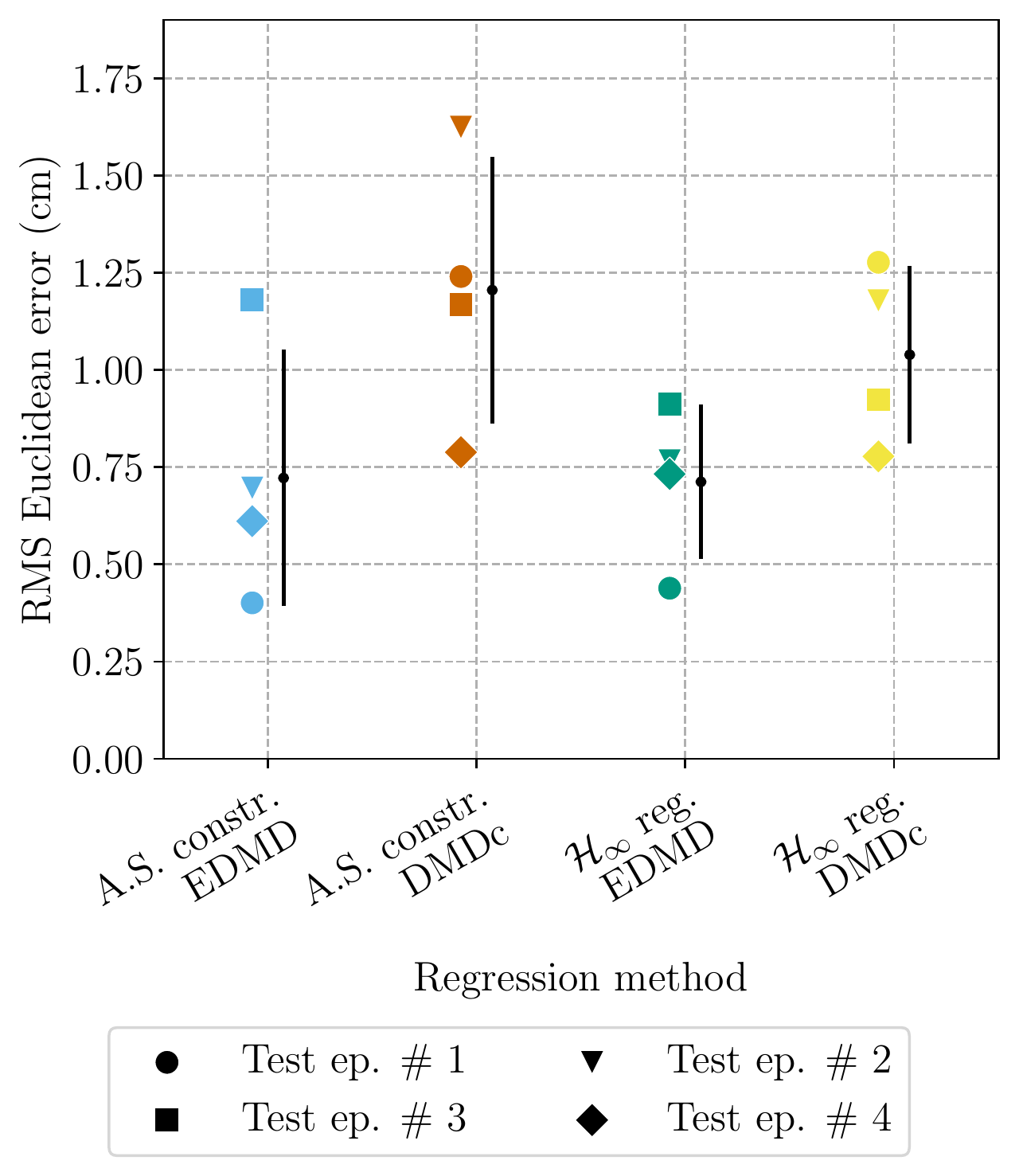}
    \end{subfigure}%
    \caption{(\subref{fig:soft_robot_dmdc_bode})~Bode plots of Koopman systems
    approximated from the soft robot dataset using EDMD and DMDc regressors.
    The DMDc methods preserve the frequency responses of the corresponding
    systems identified with the EDMD methods.
    (\subref{fig:soft_robot_scatter_dmdc})~RMS Euclidean errors of Koopman
    systems approximated from the soft robot dataset using EDMD and DMDc
    regressors. Error bars indicate standard deviation of RMS error over the
    four test episodes. Since the DMDc methods identify reduced-order
    Koopman models of the system, they have larger mean errors. However, the
    \Hinf{}~norm regularizer still significantly reduces the standard deviation.}
\end{figure}

An important decision in the DMDc algorithm is the choice of singular value
truncation method.
Optimal hard singular value truncation~\cite{gavish_2014_optimal} is used to
determine $\hat{r}$, while $\tilde{r}$ is left at full rank. For the soft robot
dataset, the optimal hard truncation algorithm retains only 14 of the 34
singular values of $\mbf{A}$.
\Cref{fig:soft_robot_dmdc_svd} demonstrates that the DMDc methods indeed
reduce the dimensionality of the problem, while also showing that the remaining
singular values are close to their EDMD counterparts.
\Cref{fig:soft_robot_dmdc_bode} shows that the frequency responses of the
original Koopman systems are preserved by the DMDc methods. In spite of their
reduced dimensionality, the Koopman systems identified with DMDc retain their
frequency domain properties.

In \cref{fig:soft_robot_scatter_dmdc}, the RMS Euclidean errors of the
EDMD and DMDc methods with asymptotic stability constraints and \Hinf{}~norm
regularization are summarized.
Since the Koopman systems identified by the DMDc methods are of a lower order,
their mean prediction error is higher than that of the EDMD methods.
However, the \Hinf{}~norm regularizer retains its benefit of improving
prediction consistency. Furthermore, as demonstrated by
\cref{fig:soft_robot_dmdc_bode}, the frequency response properties of the
EDMD regression methods are preserved by the reduced-order DMDc models.

Note that, in one case, the Koopman system identified using
stability-constrained DMDc diverged due to poor numerical conditioning. This
short segment of the dataset was omitted throughout the paper to allow for a
more fair comparison between regression methods. This finding highlights the
advantages of the \Hinf{}~regularization method in identifying numerically
well-conditioned Koopman matrices.

\begin{figure}[htbp]
    \centering
    \begin{subfigure}{0.5\textwidth}
        \centering
        \caption{}\label{fig:soft_robot_exec}
        \includegraphics[width=\linewidth]{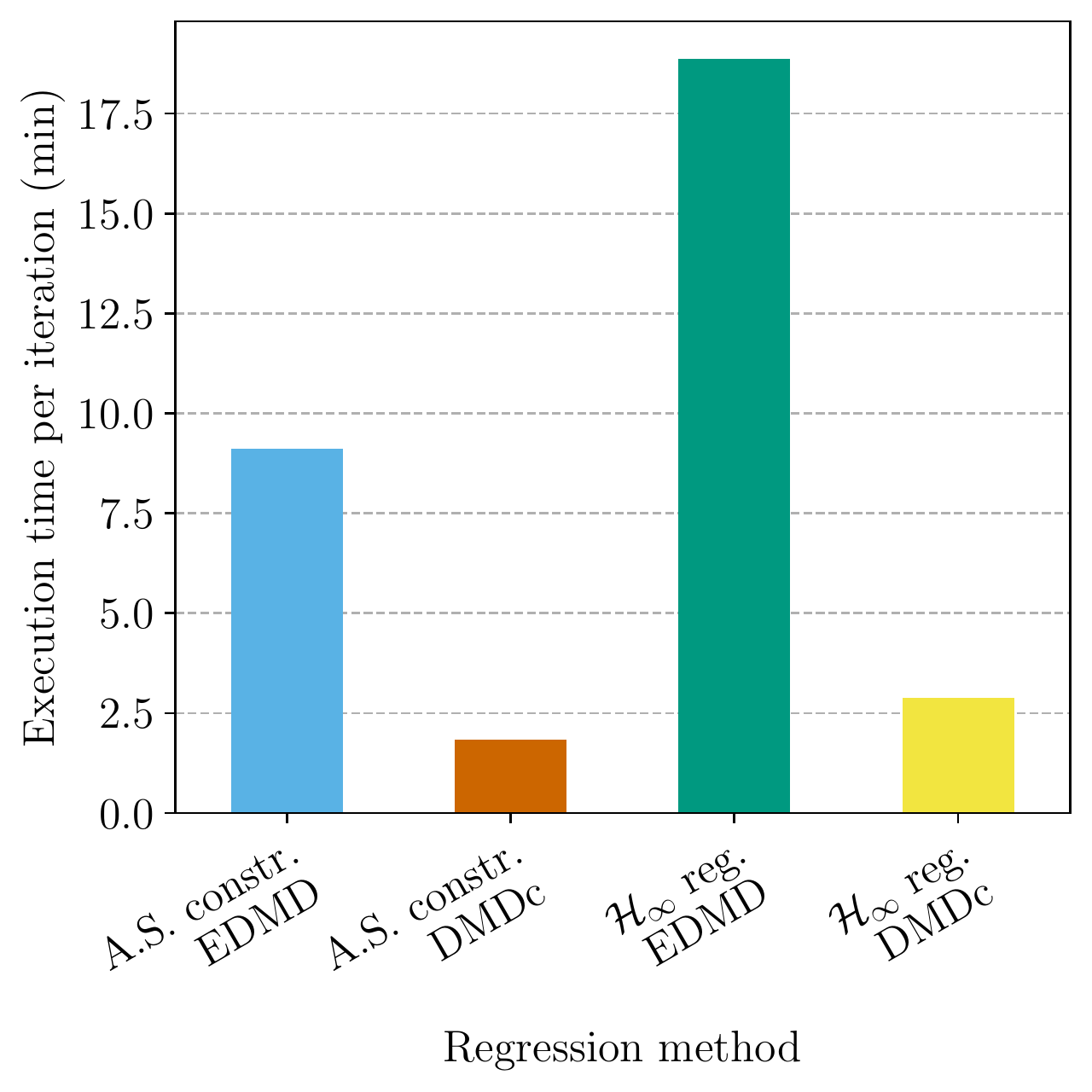}
    \end{subfigure}%
    \begin{subfigure}{0.5\textwidth}
        \centering
        \caption{}\label{fig:soft_robot_ram}
        \includegraphics[width=\linewidth]{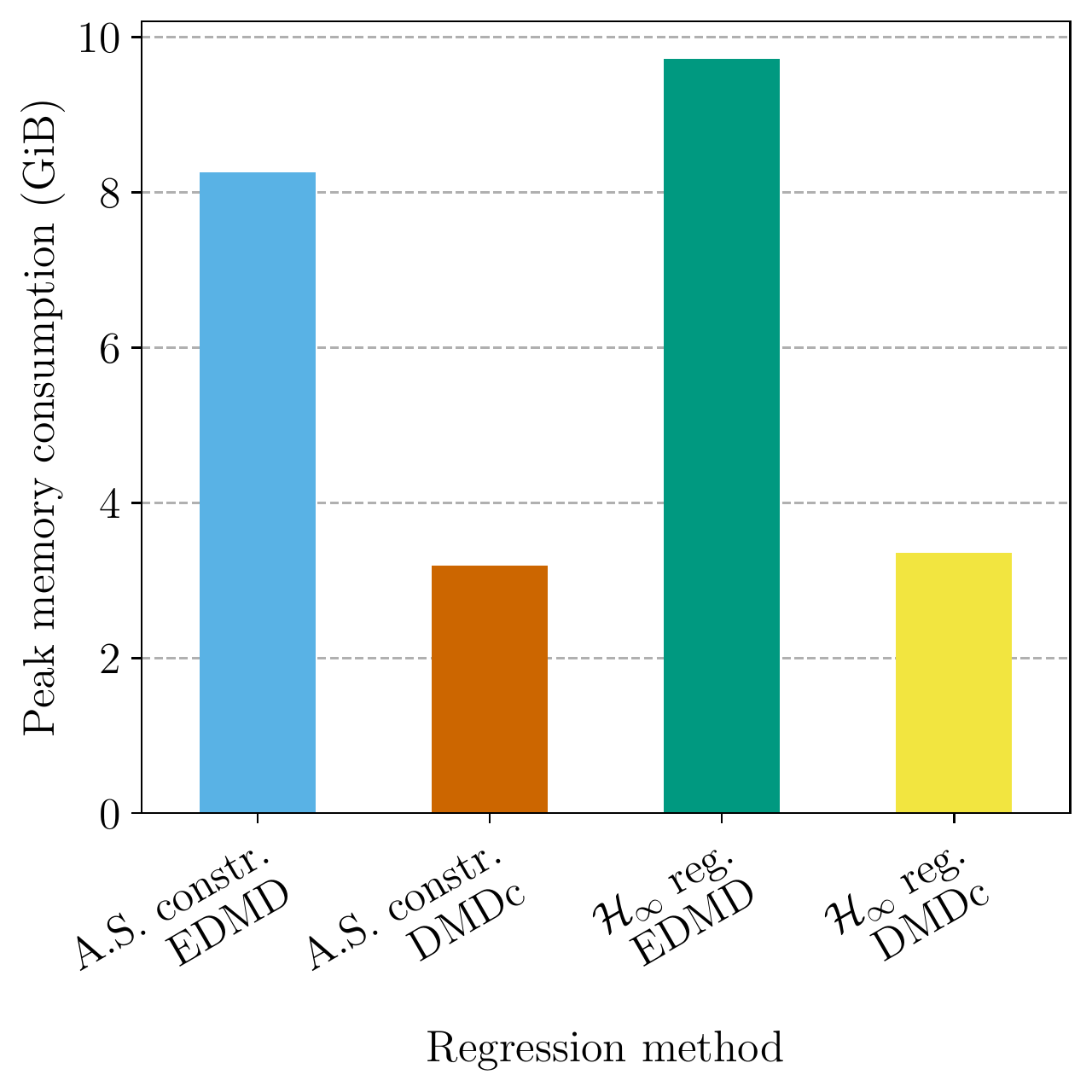}
    \end{subfigure}%
    \caption{(\subref{fig:soft_robot_exec})~Execution time per iteration and
    (\subref{fig:soft_robot_ram})~peak memory consumption of EDMD and DMDc
    regression methods using the soft robot dataset. The DMDc methods run
    significantly faster and consume less memory than the EDMD methods.
    Tests were run on a PC with an Intel Core i7-10700K processor using the
    MOSEK solver.}\label{fig:soft_robot_resources}
\end{figure}

The most important advantage of the DMDc regression methods is their
computational savings when many lifting functions are required. The long
execution times of the EDMD methods, along with their high memory consumption,
make cross-validation impractical.
\Cref{fig:soft_robot_resources} demonstrates the significant resource savings
provided by the DMDc methods in both execution time and peak memory consumption.
Memory consumption is of particular importance when running multiple instances
of a regressor in a multi-process cross-validation scheme.
The DMDc regression methods presented provide significant computational savings
while still retaining the frequency-domain characteristics of their EDMD
counterparts. In spite of their higher mean prediction error, it is often
worthwhile to leverage them for Koopman operator identification,
particularly when hyperparameter optimization is a priority.

\section{Conclusion}\label{sec:conclusion}
Approximating the Koopman matrix using linear regression proves challenging
as lifting function complexity increases. Even small problems can become
ill-conditioned when many lifting functions are required for an accurate fit.
Viewing the problem from a systems perspective, where system inputs pass through
dynamics and lead to outputs, provides multiple avenues to enforce asymptotic
stability and penalize large input-output gains in the system, thus ensuring
improved numerical conditioning.
In particular, regularizing the regression problem with the \Hinf{}~norm
provides the opportunity to tune the regularization process in the frequency
domain using weighting functions.
The significant performance savings presented by the DMDc-based regression
methods allow the \Hinf{}~norm regularizer to be applied to much larger systems
while still remaining tractable.

The nonconvex optimization problems required to use the asymptotic stability
constraint and \Hinf{}~norm regularizers limit their applicability to practical
problems.
Future research will address this limitation by making use of more efficient BMI
solution methods, including Iterative Convex
Overbounding~\cite{warner_2017_iterative} and branch-and-bound
methods~\cite{vanantwerp_2000_tutorial}.
Although the use of the \Hinf{}~norm~\cite[\S3.2]{caverly_2019_lmi} as a
regularizer is explored in this paper, any system norm, like the
\Htwo{}~norm~\cite[\S3.3]{caverly_2019_lmi} or a mixed
\Htwo{}~norm~\cite[\S3.5]{caverly_2019_lmi}, can be used. The unique properties
of system norms prove useful in addressing the numerical challenges associated
with approximating the Koopman operator from data, and will be explored further
in future work.

\dataccess{%
The methods presented in this paper and its
predecessor~\cite{dahdah_2021_linear} are implemented in release \texttt{v1.0.4}
of \texttt{pykoop}, the authors' open source Koopman operator identification
library~\cite{dahdah_2021_pykoop}.
The code required to reproduce the plots in
this paper is available in a companion repository at
\url{https://github.com/decarsg/system_norm_koopman}, release \texttt{v1.0.3}.}

\aucontribute{Both authors conceived of the research and performed the formal analysis.
S.D.\ wrote the software, performed the experiments, and prepared the visualizations.
J.R.F.\ supervised the research, formulated its high-level objectives, and assisted
with detailed derivations.}

\competing{The authors declare that they have no competing interests.}

\funding{%
This work was supported by the Mecademic Inc.\ through the Mitacs Accelerate
program, and by the Natural Sciences and Engineering Research Council of
Canada (NSERC), the National Research Council of Canada (NRC), the Toyota
Research Institute, the National Science Foundation Career Award [grant number 1751093],
and the Office of Naval Research [grant number N00014-18-1-2575].}

\ack{%
The authors thank Daniel Bruder, Xun Fu, and Ram Vasudevan for graciously
providing the soft robot dataset used in this research. The authors also
acknowledge Doug Shi-Dong, Robyn Fortune, Shaowu Pan, Karthik Duraisamy, and
Matthew M. Peet for productive discussions about regularization techniques,
the Koopman operator, and methods for handling BMI constraints.}


\bibliographystyle{RS}

\bibliography{references}

\end{document}